\documentclass[prd,aps,showpacs,superscriptaddress,nofootinbib,notitlepage,floatfix]{revtex4-2}
\usepackage{epsfig,graphicx,amsfonts,amsmath,amssymb}

\usepackage{amsmath}
\usepackage{amssymb}

\DeclareMathAlphabet{\msfsl}{OT1}{cmtt}{m}{sl}
\RequirePackage{mathrsfs}
\usepackage{color}
\usepackage{epsf}
\usepackage{epstopdf}

\def\eq#1{{Eq.~(\ref{#1})}}
\newcommand{\Le}{\left(}
\newcommand{\Ra}{\right)}

\newcommand{\beq}{\begin{equation}}
	\newcommand{\eeq}{\end{equation}}
\newcommand{\beqar}{\begin{eqnarray}}
	\newcommand{\eeqar}{\end{eqnarray}}
\newcommand{\D}{\partial}

\newcommand{\g}{{\rm g}}

\newcommand{\ep}{\varepsilon}

\newcommand{\mB}{\mathcal{B}}
\newcommand{\mA}{\mathcal{A}}

\newcommand{\tmA}{\tilde{\mathcal{A}}}
\newcommand{\tph}{\tilde{\phi}}

\begin{document}
	\title{Graviton reggeization and high energy gravitational scattering of scalar particles}
	\author{S. Bondarenko}
	\affiliation{Ariel University, Ariel 4070000, Israel}

\date{\today}
	
\begin{abstract}

 In this paper we consider a high energy scattering of free scalar particles through a gravitational field. The one particle t-channel amplitude of the scattering in this limit is governed by reggeized graviton. Therefore, we discuss an appearance of the reggeized gravitons in the framework of Einstein-Hilbert gravity and consider Lipatov's effective action for the reggeized gravitons. We calculate the trajectory of the corresponding $t$-channel amplitude in the framework and thereafter define the leading order amplitude of scattering of two massive scalar particles. 
An impact factor of the interaction of scalar particle with the reggeized gravitons is also calculated and possible applications of the approach are discussed as well.
\end{abstract}
	
	\maketitle

\section{Introduction}

 A reggeization of the scattering amplitude (propagator) in high energy scattering of different types of particles is very well known and deeply studied result in quantum field theories, 
see \cite{Regge,BFKL,BFKL1,BFKL2}. In the phenomenological models of hadron scattering, the idea that at high energies the amplitude is determined by singularities in the $t$-channel partial waves was resulted in the construction of Gribov's reggeon calculus, see \cite{Gribov}.  In QCD, in turn, the reggeization manifests itself in an appearance of new degrees of freedom, aka reggeized gluons and quarks, and in the appearance of colorless bound state of two reggeized gluons, Pomeron states, which contribution to the amplitude of inclusive cross sections is leading at a high energy limit, see for the details \cite{BFKL,BFKL1,BFKL2,LipatovEff,LipatovEff1,LipatovEff2}. The Pomeron's interactions in the theory, in turn, lead to the  construction of QCD Regge Field Theory (QCD RFT), see 
\cite{LipatovEff,LipatovEff1,LipatovEff2,Our,Our01}, which is a generalization of the Gribov's reggeon approach for the case of QCD degrees of freedom. The reggeization of amplitude with $t$-channel graviton's interactions between scattered particles was studies as well, 
see \cite{LipatovGrav,LipatovGrav01,LipatovGrav02,LipatovGrav03,LipatovGrav04,LipatovGrav05,LipatovGrav06,LipatovGrav1,LipatovDL}.

 An important issue behind all these calculations is a determination of possible unitary corrections to the scattering amplitudes. In QCD the calculations of this type can be done in the framework of the initial BFKL (Balitsky,Fadin,Kuraev,Lipatov) approach, see \cite{BFKL,BFKL1,BFKL2,Fadin,Fadin1,Fadin2,Fadin3,Fadin4,Fadin5,Fadin6} or in the framework of the Lipatov's effective action, see 
\cite{LipatovEff,LipatovEff1,LipatovEff2, LipatovGrav,LipatovGrav01,LipatovGrav02,LipatovGrav03,LipatovGrav04,LipatovGrav05,LipatovGrav06,LipatovGrav1,LipatovDL,Our,Our01,SabV1,SabV2,SabV3,SabV4,SabV5,SabV6} for the  QCD and gravity frameworks. The advantage of the use of the effective Lagrangians in these calculations is that the calculations can be formalized on the base of powerful QFT methods, 
see \cite{Our,Our01,SabV1,SabV2,SabV3,SabV4,SabV5,SabV6,Our1,Our11,Our12}. In this case we can obtain amplitudes of the scattering as whole using, for example, the \cite{Faddeev,Faddeev1} formalism of calculation of effective action, see also \cite{Our2}.
In any case, the first step in an validation of the correctness of the effective action's approaches is an calculation of the known quantities in that framework and comparison of the results with the  known answers. The quantities we calculate usually first are the Regge trajectories of the scattering amplitudes, for QCD it was done in \cite{Our,Our01} whereas for the gravity effective action of \cite{LipatovGrav1}  the task was not complete.

 The reason for lack of the direct calculations of Regge trajectory of the Einstein-Hilbert graviton in the Lipatov's effective action framework is simple. There are relations between 
QCD and gravity amplitudes at high energies, the results from QCD side can be applied in the gravity sector by use of methods and calculation techniques 
from QCD, see for example \cite{Venug,Venug1,Venug2,SabVera,SabVera1,SabVera2,SabVera3,OurZubkov}.
Nevertheless, the problems of effective action construction and the direct calculations itself in the framework of Einstein-Hilbert gravity are difficult. 
Namely, in QCD the Lipatov's effective action approach can be formulated 
as calculation of scattering amplitude of two particles at high energies when the propagators of the particles, at LO approximation, are represented by the Wilson lines 
(ordered exponentials), see \cite{Meggio,Meggio1,Meggio2,Meggio3,Meggio4,Meggio5,OurZubkov1}.  Formally defining, we construct an effective action in rapidity space which describes averaging of two interacting Wilson lines
in respect to regular QCD fields, see details of the approach in \cite{OurZubkov1}. 
The reggeized fields in this case arise in the action due a Hubbard-Stratonovich type of transformation. Performing the transform, in the final action appear two effective currents which interact with axial fields introduced by the transform, these axial fields are reggeized fields we discuss. The form of the effective currents provides a definition of the reggeized fields in the formalism as classical solutions for the corresponding regular fields in the initial QCD Lagrangian.  Integrating out the initial degrees of freedom we stay therefore with only reggeized fields present in the effective action, the resulting Lagrangian represents a corresponding RFT model. The whole scheme, in this extend, is based on the particular high energy form of the initial particle's propagators, 
see \cite{LipatovEff,LipatovEff1,LipatovEff2}.
Written as Wilson lines, they preserve the gauge symmetries of the original Lagrangian and satisfy all our field symmetry queries. An appearance of these ordered exponential is related, of course, to the gauge symmetries of the theory and, technically, to the gauge covariant derivatives in the Lagrangian. In gravity we have the similar situation in the Einstein-Cartan gravity formalism, 
see \cite{OurZubkov1}, but the Einstein-Hilbert formulation of the gravity we discuss is much more difficult to handle. Here we have no well defined notion of the ordered exponential,
see nevertheless \cite{GravExpon,GravExpon1,GravExpon2,GravExpon3,GravExpon4,GravExpon5,GravExpon6} and references therein as example of construction of  Wilson line operators in the Einstein-Hilbert formulation of gravity.

 The problem we face is simple in fact. We can define a particle's propagator in the background gravitational field, but it can not be reduced in general to some fully enclosed definition 
of propagator as an ordered exponential plus some corrections
valid in high-energy limit as it was done in QCD case\footnote{It can be done, of course, in a gauge formulation of gravity theory, as mentioned above}. 
The resolution of the problem was proposed by L.Lipatov in 
\cite{LipatovGrav1} where he proposed to construct the effective currents of the theory perturbatively, order by order, by request of diffeomorphism invariance of the obtaining expression. 
The procedure indeed allows perturbatively construct the currents, but it has its own drawbacks. First of all, it is not clear if the obtained answer is complete. Whereas in QCD we expand the final expression in order to get terms of requested perturbative order and control all the corrections, in gravity we have perturbative calculations from down to up and we can not guarantee that  we do not miss some contributions. Secondly,
the reggeization arises as consequence of the light-cone time ordering of the terms in the ordered exponentials aka effective currents. Such terms necessarily arise in the proposed approach, but the
way of the ordering of these terms is not clear in general and we have additionally to determine the ordering of product of few operator of this kind in the expressions. Of course, an additional complication is a complex structure of the gravity, see nevertheless \cite{Venug,Venug1,Venug2}. Therefore, the trajectory of the reggeized graviton, 
obtained in \cite{LipatovGrav,LipatovGrav01,LipatovGrav02,LipatovGrav03,LipatovGrav04,LipatovGrav05,LipatovGrav06}, 
was not directly recalculated in the framework of effective action of  \cite{LipatovGrav1}. This task, to our opinion, is important due the wide use of different effective gravity theories, see for example \cite{Veneziano,Veneziano1,Veneziano2,Veneziano3,Veneziano4,Veneziano5,Veneziano6} and references therein. The validation of the Lipatov's gravitational effective action, therefore, will contribute to the establishing of the approach as a tool for an explore of quantum gravity structure
at high energy limit, see
\cite{Donoghue,Donoghue1,Donoghue2,Donoghue3} for the formulation of the classical gravity as effective low energy quantum field theory.

  An another important issue can be explored through the effective action approach is an reggeization of the graviton's propagator and it's consequences. Whereas in QCD, for example, we talk about
gluons or quarks scattering and gluon's reggeization, in gravity we have any particles scattering through the reggeized gravitons. It must be underlined therefore, that in in general the $t$-channel 
amplitudes can be constructed not only from gravitons, which provide a leading contribution of course, but also from the particles participate in the scattering. In particularly, we consider the case of massive free scalar particles scattering and
here there is the off-shell $t$-channel scalar amplitude in the process of the two particle to two gravitons scattering. An another issue, it is an arising of the reggeized graviton in the framework.  
In general, in the given scheme, the reggeization appears as a consequence of the presence of special kind of energy-momentum tensor in the Einstein-Hilbert (EH) linearized gravity Lagrangian. 
An important question is 
what is the origin of this tensor there. 
As mentioned above, in the Lipatov's effective action approach such term represents an interaction between propagators of two energetic particles averaged in respect to 
quantum fields, in the HE gravity such construction is not defined yet. Therefore, we need to clarify the appearance of the reggeized graviton as consequence of some interactions between scalar particles at
high energy limit. Namely, similarly to QCD, we have there as well a propagation of high energy particles accompanied by radiation of $t$-channel soft gravitons. Additional interesting issue is that the reggeization, in general, leads to a change of the amplitude, i.e. to an arising of a Regge trajectory with corresponding slope which is important in a definition and calculation of differential, diffractive and productive amplitudes, see \cite{Ross,GriB} for the case of QCD. Of course, also, the reggeized amplitude  at whole can be important if we talk about any quantum gravity amplitude calculations.

 Therefore, the paper is organized as following. In the next Section we discuss a classical picture of gravity field produced by the high-energy massive scalar particles, the field has  
a shock-wave form.
We consider there a calculation of the corresponding energy-momentum tensor and corresponding solutions for the graviton's components.
In second Section we generalize the classical picture for the quantum  case introducing reggeized gravitons and corresponding effective currents. In this  Section \ref{Curr} we repeat 
and clarify calculations of the effective currents from the \cite{LipatovGrav1} paper introducing reggeized gravitons as new degrees of freedom in the approach. Section \ref{GravPart}
is dedicated to the general effective action which includes $t$-channel off-shell scalar particle interactions  and effective gravity counterpart, an generalization of the action for the case of
particle's production is there as well. In the following Section \ref{EffA} we discuss an issue of construction of quantum effective action and scattering amplitudes basing on the Lipatov's 
effective action approach. In the following Section \ref{ImpFac} we present calculations of impact factors of the scattering, the impact factors are introduced as parts of effective action at the edges of rapidity there. The $t$-channel scattering amplitude, i.e. propagator of reggeized gravitons, is defined and calculated in the Section \ref{ScatP}, there the answer for the graviton's trajectory is 
represented as well. In the Section \ref{LOA} we calculate the LO scattering amplitude of two scalar particles through the reggeized gravitons and compare the result with a low-energy (Newtonian)
answer, the last Section \ref{Conc} is a conclusion of the paper. All relatively long calculation and long expressions we put in the Appendix \ref{AppA}-Appendix \ref{AppD} Appendices in the end of the paper with corresponding references listed in the text.

\section{General framework: classical shock wave in flat background}\label{CompM}

  In this Section we discuss a gravitational field produced by classical high-energy field taken in the form of shock-wave. 
As fist of all, consider a general Lagrangian consists Lagrangian of free scalar field, Lagrangian of interaction of the scalar field $\phi$ with the shock wave and induced Lagrangian
which describes shock-wave field Lagrangian. We have:
\beqar\label{SW1}
L/\sqrt{-g}\,&=&\,L_{\phi}(\phi)\,+\,L_{int}(\phi,\mA)\,+\,L_{ind}(\mA)\,=\,
L_{\phi}(\phi)\,+\,\phi\,J_{eff}(g,\phi)\,\mA\,+\,L_{\mA}(\mA)\,=\,
\nonumber \\
&=&\,\frac{1}{2}\Le g^{\mu \nu}\Le \D_{\mu}\phi\Ra\,\Le\D_{\nu}\phi\Ra\,-\,m^2\phi^2\Ra\,-\,
g^{\mu \nu}\Le \D_{\mu}\phi\Ra \,\Le\D_{\nu}\mA\Ra\,+\,m^2\phi\mA\,-\,\frac{1}{2}\Le \D_{\mu}\phi\Ra \,\Le\D_{\nu}\mA\Ra\,j^{\mu \nu}\,+\,
\nonumber \\
&+&\,
g^{\mu \nu}\Le \D_{\mu}\mA\Ra\,\Le \D_{\nu}\mA\Ra\,-\,\,m^2\mA^2\,,
\eeqar
here $\mA$ is a Reggeon (shock-wave) scalar field and $j^{\mu \nu}$ is an effective current which describes an interaction of the scalar and gravitational fields at higher orders. 
For a classical shock wave we take $j^{\mu \nu}\,=\,0$ staying with the standard Lagrangian.
The interaction and induced Lagrangians in the expression are constructed by request that it provides a specific classical value for the scalar field.
Considering the initial scalar particle moving in the $x^{-}$ direction we define the classical shock wave as
\beq\label{SW2}
\mA\,=\,\mA(x^{-},x_{\bot})\,,\,\,\,\,\D_{+}\mA\,=\,0\,.
\eeq
This precise form of the shock-wave assumes that the corresponding shock-wave action is defined 
for the three dimensions. The simplest way for the reduction is to remember that in calculations we account only leading with respect to $\sqrt{s}$ terms and neglect all sub-leading corrections.
Namely, the field has the following general structure:
\beq\label{CSW20001}
\mA(x^{-},x^{+},x_{\bot})\,\rightarrow\,\mA(x^{-},x_{\bot})\,+\,\tilde{\mA}(x^{-},x^{+},x_{\bot})\,,\,\,\,\,
\tilde{\mA}(x^{-},x^{+}\,=\,0,x_{\bot})\,=\,0\,,
\eeq
see \cite{OurZubkov1,Our1,Our11,Our12}.
The simplest possible parametrization of the second term above, for example, is the following expression:
\beq\label{CSW20002}
\tilde{\mA}(x^{-},x^{+},x_{\bot})\,\rightarrow\,x^{+}\,\tilde{\mA}(x^{-},x_{\bot})\,.
\eeq
In the case of the Regge kinematic (high energy kinematic), we have for the $x^{+}$ 
\beq\label{CSW20003}
x^{+}\,\sim\,1/\sqrt{s}
\eeq
that provides the mentioned smallness of the additional term in \eq{CSW20001}. We need to regularize the $\mA$ field
in order to obtain the  three dimensional theory accounting that the field is localized around $x^{+}\,\sim\,0$. 
The simplest $\delta(x^{+})$ regularization is not very convenient for our purposes. 
The quadratic and possible greater power terms could be in the Lagrangian, moreover it is simply not precise if we do not talk
about the $\sqrt{s}\,\rightarrow\,\infty$ limit. Thereby, we regularize the $x^{+}$ behavior of our field as following:
\beq\label{CSW204}
\mA(x^{-},x^{+},x_{\bot})\,\rightarrow\,C\,e^{-B*x^{+\,2}}\,\mA(x^{-},x_{\bot})\,,\,\,\,\,\,\,C^{2}\int\,dx^{+}\,e^{-2 B*x^{+\,2}}\,=\,1\,.
\eeq
It is compatible with \eq{CSW20001} general form of the field and effectively provides the reduction of the action to the three dimensional form.

 Next, solving the classical equations of motion for the $\phi$ filed to leading order (LO) precision (with respect to the gravitational field), we obtain:
\beq\label{SW4}
\phi\,=\,\mA(x^{-},x_{\bot})
\eeq
that provides the following LO \eq{SW1} Lagrangian:
\beq\label{SW5}
L\,=\,\sqrt{-g}\Le \frac{1}{2}\, g^{\mu \nu}\Le \D_{\mu}\mA\Ra\,\Le\D_{\nu}\mA\Ra\,-\,\frac{1}{2}\, m^2\mA^2\Ra\,
\eeq
which has a regular form as requested.
Using the usual definition of the momentum-energy tensor of the problem we have:
\beq\label{SW7}
T_{\mu \nu}\,=\,\frac{2}{\sqrt{-g}}\,\Le \frac{\delta L}{\delta g^{\mu \nu}}\,-\,\D_{\rho}\frac{\delta L}{\delta g^{\mu \nu}\,_{,\rho}}\Ra
\eeq
for the $(+,-,-,-)$ metric's signature, the presence of the second term in the expression is due the possible dependence of the effective current on the derivatives of the metric. We obtain 
a regular answer correspondingly:
\beq\label{SW8}
T_{\mu \nu}\,=\,\Le \D_{\mu}\mA\Ra\,\Le\D_{\nu}\mA\Ra\,-\,\frac{1}{2}\,g_{\mu \nu}g^{\rho \sigma}\Le \D_{\rho}\mA\Ra\,\Le\D_{\sigma}\mA\Ra
\,+\,g_{\mu \nu}\frac{1}{2}\, m^2\mA^2\,.
\eeq
In order to proceed, we need to define the perturbative scheme for the further calculations. 
We consider a metric determined by a weak gravitational field $h$ above some given background metric $\eta$, which is not necessary flat:
\beq\label{SW9}
g_{\mu \nu}\,=\,\eta_{\mu \nu}\,+\,h_{\mu \nu}\,.
\eeq
There are different schemes of the gauge fixing of the $h$ perturbation, see \cite{Schwarz} for example, we choose an usual 
harmonic (de Donder) gauge:
\beq\label{SW10}
\D_{\nu}\,\Le h^{\nu}_{\mu}\,-\,\frac{1}{2}\,\delta^{\nu}_{\mu}\,h\Ra\,=\,0\,.
\eeq
So, what we have now, is a perturbation expansion of the the momentum-energy tensor:
\beqar\label{SW11}
T_{\mu \nu}\,&=&\,T_{\mu \nu}^{0}\,+\,T_{\mu \nu}^{1}\,+\,\cdots \\
T_{\mu \nu}^{ind}\,&=&\,T_{\mu \nu}^{ind\,1}\,+\,T_{\mu \nu}^{ind\,2}\,+\,\cdots\,, \label{SW11001}\\
\eeqar
where $T_{\mu \nu}^{i}\,\propto\,h^{i}$. The first perturbative order of the momentum-energy tensor, therefore, defines a leading order perturbation above the background, further terms determines corrections to the leading order result.
In general the similar expansion for the effective current holds:
\beq\label{SW12}
j\,=\,j^{2}\,+\,j^{3}\,+\,\cdots \,,
\eeq
see further.
Our aim now is a determination of the LO solution for the gravitational weak field produced by the relativistic classical shock-wave particle.

 First of all, we need to  introduce 
a free classical on-shell shock wave given by a solution of the classical equation of motion:
\beq\label{CSW1}
\Box\,\mA\,=\,0
\eeq
with
\beq\label{CSW2}
\Box\,=\,\D_{\nu}\,\Le\,\sqrt{-g}\,g^{\nu \mu}\D_{\mu}\,\Ra\,+\,m^2\,.
\eeq
There are many solutions are known for the scalar field in a gravitational background, we do not discuss them here.
What is required only is that the solution we will use will have the \eq{SW2} shock-wave form:
\beq\label{CSW3}
\mA\,=\,\mA(x^{-},x_{\bot})\,.
\eeq
For the flat background with metric
\beq\label{CSW4}
\eta_{\,\mu \nu}\, =\,\eta^{\,\mu \nu}\,= \left(
\begin{array}{cccc}
0 & 1 & 0 & 0 \\
1 & 0 & 0 & 0 \\
0 & 0 & -1 & 0 \\
0 & 0 & 0 & -1
\end{array} \right)\,\,\,\,\,\mu\,,\nu\,=\,(+,\,-,\,\bot)\,
\eeq
this field is a solution of usual Klein-Gordon equation of course
\beq\label{CSW5}
\Le \D_{\mu}\D^{\,\mu}\,+\,m^2\Ra\,\mA\,=\,\Le \D_{i}\D^{\,i}\,+\,m^2\Ra \mA\,=\,0\,,\,\,\,\,m^2 \mA\,=\,\D_{i}^2 \mA\,
\eeq
written in terms of light-cone coordinates. So, in this Section, as $\mA$ we will understand a classical solution of \eq{CSW5} which has the \eq{CSW3} form
of the coordinate dependence, see Appendix \ref{AppA}. Therefore, determining the momentum-energy tensor \eq{SW8} for the \eq{CSW5} on-shell fields, we obtain to the leading, i.e $h^{0}$, order:
\beqar\label{CSW6}
T_{\mu \nu}^{\,0}\,&=&\,\Le \D_{\mu}\mA\Ra\,\Le\D_{\nu}\mA\Ra\,+\,\frac{1}{2}\,\eta_{\mu \nu}\, \D_{i}\Le\mA\,\D_{i}\mA\Ra\,;\\
T^{\,0}\,&=&\,\Le \D_{i}\mA\Ra\,\Le\D^{i}\mA\Ra\,-\,2\,\D_{i}\Le\mA\,\D^{i}\mA\Ra\,\label{CSW601} .
\eeqar
Our next step is a solution of Einstein equations. We solve them following to the \eq{SW10} gauge condition and assumption that
\beq\label{CSW7}
h\,\propto\,A(x^{-},x_{\bot})\,,\,\,\,h_{\mu \nu}=h_{\mu \nu}(x^{-},x_{\bot})\,,\,\,\,\D_{+}h_{\mu \nu}\,=\,0\,.
\eeq
The Einstein equation we obtain, have in a simple form therefore:
\beq\label{CSW7001}
\Box h_{\mu \nu}\,=\, \D_{i} \D^{\,i} h_{\mu \nu}\,=\,-\,\D_{i}^{2} h_{\mu \nu}    \,=\,-2\,\kappa\,\Le T_{\mu \nu}\,-\,\g_{\mu \nu}\frac{1}{2}\,T\Ra\,.
\eeq
Now let's consider the $R_{++}$ and $R_{+i}$ Ricci tensor components of the tensor, here we have the trivially satisfied equations:
\beqar\label{CSW8}
\D_{i} \D^{\,i} h_{++}\,& = &\,-2\,\kappa\Le T_{++}^{0}\,-\,\frac{1}{2}\,\eta_{++}\,T^{0}\Ra\,=\,0\,,\,\,\,T_{++}^{0}\,=\,\eta_{++}\,=\,0\,,\,\,\,
\D_{i}^{2} h_{++}\,=\,0\,;
\\
\D_{i} \D^{\,i} h_{+i}\,& = &\,-2\,\kappa\Le T_{+i}^{0}\,-\,\frac{1}{2}\,\eta_{+i}\,T^{0}\Ra\,=\,0\,,\,\,\,T_{+i}^{0}\,=\,\eta_{+i}\,=\,0\,,\,\,\,
\D_{i}^{2} h_{+i}\,=\,0\,\label{CSW9}.
\eeqar
Next we write the parts of the equation for the $R_{+-}$ tensor's component. We have:
\beq\label{CSW10}
T^{0}_{+-}\,-\,\frac{1}{2}\,\eta_{+-}\,T^{0}\,= \,-\,\frac{1}{2}\,\eta_{+-}\,\Le \D_{i}\mA\Ra\,\Le\D^{i}\mA\Ra\,+\,
\frac{1}{2}\,\eta_{+-}\,\D_{i}\,\Le \mA\,\D^{i}\,\mA\Ra\,=\,-\,\frac{1}{2}\,\eta_{+-}\,m^{2}\,\mA^{2}\,
\eeq
and correspondingly
\beq\label{CSW11}
\D_{i}^{2} h_{+-}\,=\,-\,\eta_{+-}\,m^{2}\kappa\mA^{2}\,.
\eeq
Our next equation is an equation for the $R_{--}$ component of the Ricci tensor. We have correspondingly:
\beq\label{CSW12}
\D_{i}^{2} h_{--}\,=\,2\kappa\,\Le \D_{-}\mA \Ra\,\Le \D_{-}\mA \Ra\,.
\eeq
For the tensor's components with the transverse indexes we have:
\beq\label{CSW13}
\D_{k}^{2} h_{- i}\,=\,2\kappa\,\Le \D_{-}\mA \Ra\,\Le \D_{i}\mA \Ra\,\,.
\eeq
and 
\beq\label{CSW13001}
\D_{k}^{2} h_{i j}\,=\,2\kappa\,\Le \Le \D_{i}\mA \Ra\,\Le \D_{j}\mA \Ra\,-\,\eta_{ij}\frac{1}{2} m^2 \mA^2 \Ra\,.
\eeq
We find, therefore, all the components of the $h_{\mu  \nu}$ tensors, so let's discuss their structure and asymptotic behavior of these fields.

 First of all, we determine the trivial components of the gravity field:
\beq\label{CSW14}
h_{+ +}\,=\,h_{i +}\,=\,0\,
\eeq
which are definitely solutions of \eq{CSW8}-\eq{CSW9}. This particular form of the solutions is consistent with the known asymptotic weak field solutions, see \cite{LipatovGrav1}. 
An additional gauge condition we have is an relation between the $h$ tensor components 
\beq\label{CSW14001}
\D_{-}h_{+-}\,-\,\D_{i}h_{i -}\,=\,\frac{1}{2} \eta_{- +} \D_{-} h\,\rightarrow\,\D_{-} h_{\bot}\,=\,\eta^{ij}\D_{-} h_{i j}\,=\,-2\, \D_{i} h_{i -}\,
\eeq
satisfied for the on-shell classical $\mA$ field.
The components of the field we have are the following therefore:
\beqar\label{CSW15}
h_{+-}(x^{-}, x_{\bot})\,& = &\,-\,\eta_{+-}\,m^{2}\kappa\,\Box^{-1}(x_{\bot},y_{\bot})\mA^{2}(x^{-},y_{\bot})\,;
\\
h_{--}(x^{-}, x_{\bot})\,& = &\,2\kappa\,\Box^{-1}(x_{\bot},y_{\bot})\, \Le \D_{-}\mA(x^{-},y_{\bot})\Ra^{2}\,;
\label{CSW16}
\\
h_{-i}(x^{-}, x_{\bot})\,& = &\,2\kappa\,\Box^{-1}(x_{\bot},y_{\bot})\, \Le \D_{-}\mA(x^{-},y_{\bot})\, \D_{i}\mA(x^{-},y_{\bot})\Ra\,;
\label{CSW16001}
\\
h_{i j}(x^{-}, x_{\bot})\,& = &\,2\kappa\,\Box^{-1}(x_{\bot},y_{\bot})\,\Le \Le \D_{i}\mA(x^{-},y_{\bot}) \Ra\,
\Le \D_{j}\mA(x^{-},y_{\bot}) \Ra\,-\,\eta_{ij} \frac{1}{2} m^2 \mA^2(x^{-},y_{\bot}) \Ra\,.
\label{CSW17}
\eeqar
Our field of interest is determining by the \eq{CSW5}. Therefore, if it can be written particularly as a factorized product of two functions:
\beq\label{CSW18}
\mA(x^{-},x_{\bot})\,=\,Z(x^{-})\,\Phi(x^{\bot})\,,
\eeq
we can write the gravitational weak field in the following form:
\beqar\label{CSW19}
h_{+-}(x^{-}, x_{\bot})\,& = &\,-\eta_{+-}\,m^{2}\kappa\,Z^{2}(x^{-})\Le \Box^{-1}(x_{\bot},y_{\bot})\,\Phi^{2}(y_{\bot})\Ra ;
\\
h_{--}(x^{-}, x_{\bot})\,& = &\,2\kappa\,\Le\D_{-} Z(x^{-})\Ra^{2}\,
\Le \Box^{-1}(x_{\bot},y_{\bot})\,\Phi^{2}(y_{\bot})\Ra ;
\label{CSW20}
\\
h_{-i}(x^{-}, x_{\bot})\,& = &\,2\kappa\,\Le Z(x^{-}) \D_{-} Z(x^{-})\Ra\,
\Le \Box^{-1}(x_{\bot},y_{\bot})\,\Phi(y_{\bot})\D_{i}\Phi(y_{\bot})\Ra ;
\label{CSW201}
\\
h_{i j}(x^{-}, x_{\bot})\,& = &\,2\kappa\,Z^{2}(x^{-})\Le \Box^{-1}(x_{\bot},y_{\bot})\,
\Le \D_{i}\Phi(y_{\bot}) \, \D_{j}\Phi(y_{\bot}) \,-\,\eta_{ij} \frac{1}{2} m^2 \Phi^2(y_{\bot}) \Ra\Ra .
\label{CSW21}
\eeqar
Now we note that 
\beq\label{CSW22}
\frac{\D}{\D x^{i}}\,=\,\frac{x_{i}}{r^{1/2}}\,\frac{\D}{\D r}\,,\,\,\,r\,=\,\sqrt{-x^{i}\,x_{i}}\,.
\eeq
Therefore, at the limit of $r\,\rightarrow\,\infty$, the leading terms are which have less spatial derivatives. 
This statement can be checked directly with the help of more general solution of 
\eq{CSW5} represented in Appendix \ref{AppA}.The second observation is that
\beq\label{CSW23}
\D_{-}\mA\,=\,-\imath\,\int_{0}^{\infty}\,dk^{+}\,k^{+}\,\tilde{\mA}(k^{+})\,e^{-\imath x^{-} k_{-}}\,,
\eeq
i.e. terms with $\D_{-}$ derivative are less suppressed in an integration over large values of $k^{+}$.
Therefore, the general statement is that we have the following hierarchy of the non-zero weak field components: 
\beq\label{CSW24}
h_{--}(x^{-}, x_{\bot})\,>\,h_{-i}(x^{-}, x_{\bot})\,>\,\Le h_{-+}(x^{-}, x_{\bot})\,\sim\,h_{i j}(x^{-}, x_{\bot})\Ra\,,
\eeq
which satisfies the high energy limit constraint
\beq\label{CSW2401}
k_{-}^{2}\,\gg\,m^2
\eeq
of course.
We see, that we face the well known result, the terms with maximum degree of $\D_{-}$ derivative in the Lagrangian are leading in the high energy limit.
Here we accounted also that  
\beq\label{CSW25}
h_{i j}(x^{-}, x_{\bot})\,\propto\,-\,2\kappa\,\eta_{ij} m^2\,Z^{2}(x^{-})\Le \Box^{-1}(x_{\bot},y_{\bot})\,\Phi^2(y_{\bot}) \Ra\,=\,
2\,\eta_{i j}\,\eta^{+ -}\,h_{+ -}\,.
\eeq
at asymptotically large $r$.

\section{Gravitational Reggeon fields and effective currents}\label{Curr}

 Our next step is an application of the of the obtained results for the construction of the gravitational Regge field theory (RFT) following the ideology of \cite{LipatovGrav1}. 
First of all, we note that the part of the \eq{SW5} Lagrangian of the scalar field which describes it interaction with the weak field gravity can be rewritten as
\beq\label{CE1001}
L\,=\,\frac{1}{2\kappa}\,\D_{i}^{2} \mB_{\mu \nu}(A)\,\Le h^{\mu \nu}\,+\,j^{\,\mu \nu}(h)\Ra\,.
\eeq
We do not write here the $\sqrt{-g}$ term, assuming it's expansion, the $\mB_{\mu \nu}(A)$ field in turn must to provide a correct leading order answer for the 
$h_{\mu \nu}$ fields which was calculated in the previous Section. The LO equation of motion of $h_{\mu \nu}$ is trivial:
\beq\label{CE1002}
h_{\mu \nu}\,=\,\mB_{\mu \nu}\,,\,\,\,\,\mB_{\mu \nu}\,\propto\,\kappa\,\D_{\mu} \mA\,\D_{\nu} \mA\,.
\eeq
In the case of classical process, the introduced $\mB$ field, which represents the classical solution of the gravitational weak field, can be written in terms of classical shock wave 
through \eq{CSW19}-\eq{CSW21} and can be understood as some new degree of freedom arising in our problem.

  Now consider a case of the scattering of relativistic scalar particles where we construct the off-shell t-channel gravitational Reggeon $B$ field which structure is 
the same as constructed from the \eq{CE1002} $t$-channel off-shell scalar field $\mA$.
Requesting that the \eq{CE1002} solution will appear as well  and using the \eq{CE1001} Lagrangian together with the gravity Lagrangian $L_{EH}$ and induced Lagrangian $L_{ind}$, we can write the general Lagrangian for the scattering problem:
\beq\label{CE100101}
L\,=\,L_{EH}(\eta,h)\,+\,\frac{1}{2\kappa}\,\D_{i}^{2} \mB_{\mu \nu}\,\Le h^{\mu \nu}\,+\,j^{\,\mu \nu}(h)\Ra\,+\,L_{ind}(\mB)\,.
\eeq
Solving the equation of motion and obtaining $\eq{CE1002}$ solution we have correspondingly:
\beq\label{CE100102}
L\,=\,L_{EH}(\eta,\mB)\,+\,\frac{1}{2\kappa}\,\D_{i}^{2} \mB_{\mu \nu}\,\Le \mB^{\mu \nu}\,+\,j^{\,\mu \nu}(\mB)\Ra\,+\,L_{ind}(\mB)\,.
\eeq
We see now that the kinematic term for the $\mB$ field is present in the general Lagrangian twice,  through $L_{EH}(\eta,B)$ and in the second term as well.
Namely, we have here a doubling of the kinetic term problem, see discussion in \cite{OurZubkov1}. Therefore, the induced part of the Lagrangian must be constructed 
in the form to cancel
this extra $\mB$ field kinetic term, we have finally:
\beq\label{CE100103}
L\,=\,L_{EH}(\eta,h)\,+\,\frac{1}{2\kappa}\,\D_{i}^{2} \mB_{\mu \nu}\,\Le h^{\mu \nu}\,+\,j^{\,\mu \nu}(h)\Ra\,-\,\frac{1}{2\kappa}\,\mB^{\mu \nu}\,\D_{i}^{2} \mB_{\mu \nu}\,.
\eeq
We will discuss the use of the Lagrangian in more details in the next Section, now we have to determine the form of the $j$ induced curent we need for the further calculations.

  The components of the effective current we
consider are the mostly contributing components of the $\mB_{\mu \nu}$ field, see above and in \cite{LipatovGrav1}. The Lagrangian of the interaction of gravitational Reggeons with the gravity 
has the following form therefore:
\beq\label{CE1008}
L_{int}\,=\,\frac{1}{2\kappa}\,\D_{i}^{2} \mB_{- -}\,\Le h^{- -}\,+\,j^{\,- -}(h)\Ra\,+\,\frac{1}{2\kappa}\,
\D_{i}^{2} \mB_{- i}\,\Le h^{- i}\,+\,j^{\,- i}(h)\Ra\,.
\eeq
Here the expression for only one from the pair of the currents is presented of course, i.e.
for the sake of shortness of notations, we consider the presence of
only $h_{++}$ and $h_{+i}$ fields in the expressions, we also assume dependence of the current  on independent field  $h_{+i}$  only and not on $h_{i+}$.
These Reggeon fields satisfies as well the kinematic constraint
\beq\label{CE1009}
\D_{+}\mB_{\mu \nu}\,=\,0
\eeq
that provides
\beq\label{CE100901}
\D_{+}h_{- -}\,=\,\D_{+}h_{- i}\,=\,0
\eeq
as well. 
In the C.M. frame of system of the particle and gravitational field, at high energy, the $h^{- i}$ and $h^{--}$ in turn satisfy
\beq\label{CE101} 
\D_{-}h^{- i}\,=\,\D_{-}h^{- -}\,=\,0\,
\eeq
kinematic constraints, see \cite{OurZubkov1} for the detailed explanation of the RFT (Regge Field Theory) construction.
Following to \cite{LipatovGrav1}, we proceed the calculations by requiring of the invariance of the effective current in respect to diffeomorphisms:
\beq\label{CE1}
\D_{i}^{2}\mB_{\mu \nu}\,\frac{\delta j^{\,\mu \nu}}{\delta h^{\rho \sigma}}\,\delta h^{\rho \sigma}\,=\,
\D_{i}^{2}\mB_{\mu \nu}\,\frac{\delta j^{\,\mu \nu}}{\delta h_{\rho \sigma}}\,\Le \varepsilon_{\rho;\,\sigma}\,+\,\varepsilon_{\sigma;\,\rho}\Ra\,=\,0\,
\eeq
with $\varepsilon$ as an infinitesimal diffeomorphism parameters. We have to consider the equations with respect to the different $\ep_{\mu}$ 
combining together terms of the same order of $h$.
\begin{enumerate}
\item
 The first equality we consider is for the $\ep_{+}$ parameter. 
We have:
\beq\label{CE3}
2\D_{i}^{2}\mB_{\mu \nu}\,\frac{\delta j^{\,1\,\mu \nu}}{\delta h_{++}}\,\D_{+} \ep_{+}\,+\,
\D_{i}^{2}\mB_{\mu \nu}\,\frac{\delta j^{\,1\,\mu \nu}}{\delta h_{+i}}\,\D_{i} \ep_{+}\,-\,
2\D_{i}^{2}\mB_{\mu \nu}\,\frac{\delta j^{\,0\,\mu \nu}}{\delta h_{++}}\,\Gamma^{+}_{++}\ep_{+}\,-\,
2\D_{i}^{2}\mB_{\mu \nu}\,\frac{\delta j^{\,0\,\mu \nu}}{\delta h_{+i}}\,\Gamma^{+}_{+i}\ep_{+}\,
=\,0\,.
\eeq
The Christoffel symbols in the expression are zero because the $\D_{\pm} h_{\mp \mu}\,=\,0$ constraints
and the equation we have is the following one:
\beq\label{CE5}
\D_{i}^{2}\mB_{--}\,\Le 2\frac{\delta j^{1\,--}}{\delta h_{+ +}}\,\D_{+} \ep_{+}\,+\,
\frac{\delta j^{1\,--}}{\delta h_{+i}}\,\D_{i} \ep_{+}\Ra\,+
\D_{i}^{2}\mB_{-i}\,\Le 2\frac{\delta j^{1\,-i}}{\delta h_{+ +}}\,\D_{+} \ep_{+}\,+\,
\frac{\delta j^{1\,-i}}{\delta h_{+j}}\,\D_{j} \ep_{+}\Ra\,=\,0\,.
\eeq
The way of solution of the equation is borrowed from the \cite{LipatovGrav1} paper. 
We can write the each term in \eq{CE5} in terms of some new variables, i.e. for example
\beq\label{CE501}
\D_{i}^{2}\mB_{--}\,\Le 2\frac{\delta j^{1\,--}}{\delta h_{+ +}}\,\D_{+} \ep_{+}\,+\,
\frac{\delta j^{1\,--}}{\delta h_{+i}}\,\D_{i} \ep_{+}\Ra\,=\,
\D_{i}^{2}\mB_{--}\,\frac{\delta j^{1\,--}}{\delta X}\,\Le 2\frac{\delta X}{\delta h_{+ +}}\,\D_{+} \ep_{+}\,+\,
\frac{\delta X}{\delta h_{+i}}\,\D_{i} \ep_{+}\Ra\,.
\eeq
Next we define new variables which satisfy the both terms in \eq{CE5} through the \eq{CE501} identity:
\beqar\label{CE6}
X_{+i}\,&=&\,\D_{+}^{-1}\,\Le \D_{+}\,h_{+i}\,-\,\frac{1}{2}\,\D_{i} h_{++} \Ra\, ; 
\\
Y_{i j}\,&=&\,A\,\D_{+}^{-1}\Le \Le \D_{i} h_{+j}\,+\,\D_{j} h_{+i}\Ra\,-\,\D_{+}^{-1}\D_{i}\D_{j} h_{++}\Ra\,=\,
\D_{+}^{-1} \Le \D_{i}X_{+j}\,+\,\D_{j}X_{+i} \Ra\,;
\label{CE7}\\
Z_{i j}\,&=&\,\frac{1}{2}\,\D_{+}^{-1} \Le \D_{i}h_{+j}\,-\,\D_{j}h_{+i} \Ra\,=\,\frac{1}{2}\D_{+}^{-1} \Le \D_{i}X_{+j}\,-\,\D_{j}X_{+i} \Ra\,.
\label{CE8}
\eeqar
We note that
whereas the first two variables are solution of both parts of \eq{CE5} by construction, the third variable does not depend on $h_{++}$ and satisfies the 
\eq{CE5} through it's following identity:
\beq\label{CE801}
\frac{\delta Z_{ij}}{\delta h_{+ k}}\,\D_{k}\ep_{+}\,=\,\frac{1}{2}\,\D_{+}^{-1}\,\Le \D_{i}\delta^{k}_{j}\,-\,\D_{j}\delta^{k}_{i}\Ra\,\D_{k}\ep_{+}\,=\,
\frac{1}{2}\,\D_{+}^{-1}\,\Le \D_{i}\D_{j}\,-\,\D_{j}\D_{i}\Ra\,\ep_{+}\,=\,0\,.
\eeq
Correspondingly, as the solution of the \eq{CE15}, we can consider effective currents built from the variables above:
\beqar\label{CE11}
j^{1\,--}\,&=&\,j_{++}^{1}\,\propto\,X_{+ i}\,X_{+}\,^{i}\,\propto\,X_{+ i}\,X_{+ i};
\\
j^{1\,-i}\,&=&\,-\,j_{+i}^{1}\,=\,X_{+ j}\,\Le Y^{j i}\,+\,B\,Z^{j i}\Ra\,=\,X_{+ j}\,\Le Y_{j i}\,+\,B\,Z_{j i}\Ra\,
\label{CE12} 
\eeqar
with $B$ as some additional constant. These forms of the currents are dictated by request that the currents are quadratic functions of $h$ field, the answers are  not final of course and
can be changed when we will consider the additional constraints.

\item
 Next equation we write is for the $\ep_{-}$ parameter. Here the situation is simple, we have no $\delta\,/\,\delta h_{- \mu}$ variations for the current because
it does not depend on the $h_{- \mu}$ (see further nevertheless). Therefore the equation with $\ep_{-}$ parameter variation does not appear, i.e. it is identically zero.

\item
 The last equation we have is for the variation with respect to $\ep_{i}$. The constraint we consider is 
\beq\label{CE13}
\D_{i}^{2}\mB_{- -}\,\Le \frac{\delta j^{\,1\,--}}{\delta h_{+i}}\,\D_{+} \ep_{i}\,-\,
2\,\frac{\delta j^{\,0\,--}}{\delta h_{++}}\,\Gamma^{i}_{++}\ep_{i}\Ra\,+\,
\D_{i}^{2}\mB_{- i}\,\Le \frac{\delta j^{\,1\,-i}}{\delta h_{+j}}\,\D_{+} \ep_{j}\,-\,
2\,\frac{\delta j^{\,0\,-i}}{\delta h_{+j}}\,\Gamma^{k}_{+j}\ep_{k}\Ra\,=\,0\,.
\eeq
We have here:
\beqar\label{CE14}
\Gamma^{i}_{++}\,&=&\,\frac{1}{2}\,\eta^{i j}\Le \D_{+}h_{+j}\,+\,\D_{+}h_{+j}\,-\,\D_{j}h_{++}\Ra\,=\,-\,\D_{+}X_{+i}\,
\\
\Gamma^{k}_{+j}\,&=&\,\frac{1}{2}\,\eta^{k s}\Le \D_{j} h_{+s}\,-\,\D_{s} h_{+j}\Ra\,=\,\frac{1}{2}\,\Le \D_{k} h_{+j}\,-\,\D_{j} h_{+k}\Ra\,=\,\D_{+}\,Z_{k j}\,.
\label{CE15}
\eeqar
For the first term of the \eq{CE13} we obtain:
\beq\label{CE15001}
\frac{\delta j^{\,1\,--}}{\delta X_{+ j}}\,\frac{\delta X_{+j}}{\delta h_{+i}}\,\D_{+}\ep_{i}\,+\,2\,\Le \D_{+}X_{+i}\Ra\,\ep_{i}\,=\,0\,,
\eeq
that after accounting \eq{CE1009} condition and by integration by parts of the second term gives:
\beq\label{CE16}
\frac{\delta j^{\,1\,--}}{\delta X_{+ j}}\,\delta^{i}_{j}\,\D_{+}\ep_{i}\,=\,2\,\Le X_{+i}\Ra\,\D_{+}\ep_{i}\,\,\rightarrow\,\,
\frac{\,\delta j^{--}}{\delta X_{+ i}}\,=\,2\,\Le X_{+i}\Ra\,
\eeq
that justifies, for a first sight, the \eq{CE11} as a correct expression for the current's NLO form. Nevertheless, there is some uncertainty presents in the answer. We have there
the $\D_{+}^{-1}$ operator squared and the possible structure of this squared operator is not really defined by the \eq{CE15}-\eq{CE16}. Namely, there we have only a variation of the expression  
of some order of the effective current which does not clarify the $x^{+}$ coordinate dependence of the current because initialy it is written in terms of the metric's fields. 
Therefore, we consider the \eq{CE11} asa formal answer which precise structure will be clarified further, in the Appendix ~\ref{AppC}.
\,\newline 
  
	The second part of \eq{CE13} has the following form in turn:
\beq\label{CE17}
\frac{\delta j^{\,1\,-i}}{\delta X_{+ j}}\,\D_{+}\ep_{j}\,-\,2\,\Le \D_{+}Z_{j i}\Ra\,\ep_{j}\,=\,0\,
\eeq
or
\beq\label{CE18}
\frac{\delta j^{\,1}_{+i}}{\delta X_{+ j}}\,=\,2\,Z_{j i}\,=\,\D_{+}^{-1}\,\Le \D_{j} X_{+i}\,-\,\D_{i} X_{+j}\Ra\,.
\eeq
Unfortunate, we can  not solve this system of equations directly, but fortunately we do not need to. 
The r.h.s. of the equation is zero to the leading order of large $r$, see \eq{CSW201} amd \eq{V8} expression. To the next order with respect to large $r$, the expression is not zero but must be invariant when the 
improved Landau-Lifshitz momentum-energy pseudotensor is introduced. It is well know fact of course, see \cite{Landau}, that
\beq\label{CE19}
h_{+ i}\,\sim\,M_{i j}\,\frac{n_{j}}{r^2}
\eeq
where $M_{i j}$ is a tensor of the full momentum of the system, $n_{j}$ is a unit tensor in $j$ direction and the expression is invariant with respect to the 
rotations in the plane orthogonal to $+$ direction.
Therefore, we can consider the r.h.s. of \eq{CE18} as a constant one with respect to the $X_{+i}$ obtaining the solution of the equation in the following form:
\beq\label{CE20} 
j^{\,1}_{+i}\,=\,\epsilon_{i j}\,\epsilon_{+ k t}\,X_{+ j}\,\D_{+}^{-1}\,\D_{t} X_{+k}\,
\eeq
with $\epsilon_{i j}$ and $\epsilon_{+ i j}$ as fully antisymmetric unit tensors. In any case, further we will use only the $j^{--}$ and $j^{++}$ leading order currents in construction of the effective 
gravity Lagrangian, it is enough for our current purposes.

\end{enumerate}

 Now we can finally determine the form of the effective currents. For our purposes, see further, it is enough to consider the only quadratic term of the effective current. This term modifies the propagators of the reggeized gravitons and provides the propagator's reggeization. So, we take \eq{CE11} and \eq{CE20}
currents 
\beqar\label{CE2001}
j^{--}\,&=&\,j_{++}\,=\,-\,X_{+ i}\,X_{+}\,^{i}\,=\,X_{+ i}\,X_{+ i};
\\
j_{+i}\,&=&\,\epsilon_{i j}\,\epsilon_{+ k t}\,X_{+ j}\,\D_{+}^{-1}\,\D_{t} X_{+k}\,
\label{CE2002}
\eeqar
as final answers for the \eq{CE1001} Lagrangian, see \cite{LipatovGrav1} for the answer with higher orders of the currents calculated.

 Yet, we put attention, that there is some higher order contribution which was not considered in \cite{LipatovGrav1} paper. Namely, we can add to the
full efective current some new term of an unusual form:
\beq\label{CE2003}
j^{--}\,=\,j_{++}\,=\,\Le  h_{++}\,+\,X_{+ i}\,X_{+ i}\Ra\,\Le 1\,-\,\gamma\,\D_{j}\D_{-}^{-1} X_{-j}\Ra\,,
\eeq
the $\gamma$ here is a constant which precise value we will discuss below. The form of the current is unusual because it mixes the two types of the Reggeon fields. Nevertheless, 
we see that the new \eq{CE2001} current does satisfy the \eq{CE5}  constraint and 
it also satisfies, by construction, the constraint which could obtained by variation  of the current with respect to the
$\ep_{-}$ parameter.  The variation with respect $\ep_{i}$ and \eq{CE13} is more tricky of course, but it will provide the term of the third order with respect to the field in the effective current,
we will not consider that precision here postponing the calculations for an additional paper. About the $\gamma$ constant we note the following, let's compare the following two terms from the both effective currents:
\beqar
\,j_{++}\,&\sim &\,-h_{++}\,\Le h_{+i}\,\D_{i}\,\D_{+}^{-1}\,+\,\gamma\D_{-}^{-1}\,\D_{j}\,h_{-j}\Ra\,\sim\,h_{++}\,C_{1};
\nonumber \\
\,j_{--}\,&\sim &\,-h_{--}\,\Le h_{-i}\,\D_{i}\,\D_{-}^{-1}\,+\,\gamma\D_{+}^{-1}\,\D_{j}\,h_{+j}\Ra\,\sim\,h_{--}\,C_{2}\,.
\eeqar
If we request for the $C_{1,2}$ functional constants (i.e. some functions which have no Riemann indexes) that $C_{1}\,=\,C_{2}$ then we obtain immediately that $\gamma\,=\,1$, we discuss the issue later as well.
These new terms in the currents provide us with the higher order terms in respect to the weak field. Further we will not consider their contributions
except the calculation of the propagator of reggeized gravitons in Section \ref{ScatP} where it can be important and where we will discuss the sign and value of the $\gamma$ parameter again.
The order of the terms in the currents we keep is the third one (fourth in the action), we do not discuss the fourth order terms in the currents
because clearly the \eq{CE2003} does not provides all of them. 
We note additionally, that, perhaps, this task of the all orders effective currents construction is easier to perform in the framework of the Einstein-Cartan gravity formalism where we can operate with a well defined Wilson line operators, see \cite{OurZubkov,OurZubkov1}.

 The last but not least remark is about the \eq{SW10} gauge condition we applied. In the given set-up we can define:
\beqar\label{CE21}
h_{+ -}\,& = &\,0\,;
\\
\D_{i} h_{i+}\,& = &\,-\,\frac{1}{2}\,\D_{+}\Le \eta^{i j}h_{ij}\Ra\,;
\label{CE22} \\
\D_{i} h_{i-}\,& = &\,-\,\frac{1}{2}\,\D_{-}\Le \eta^{i j}h_{ij}\Ra\,;
\label{CE23} \\
\D_{i} h_{ij}\,& = &\,-\,\frac{1}{2}\,\D_{j}\Le \eta^{i j}h_{ij}\Ra\,,
\label{CE24} 
\eeqar
that in turn provides
\beqar\label{CE25}
\eta^{i j}h_{ij}\,& = &\,-\,\Le \D_{+}^{-1}\D_{k} h_{k+}\,+\,\D_{-}^{-1}\D_{k} h_{k-}\,\Ra\,;
\\
\D_{i} h_{ij}\,& = &\,\frac{1}{2}\,\D_{j}\,\Le \D_{+}^{-1}\D_{k} h_{k+}\,+\,\D_{-}^{-1}\D_{k} h_{k-}\,\Ra\,.
\label{CE26}
\eeqar
Solution of  \eq{CE26} plus given \eq{CE21} component will determine the all metric tensor components in terms of the 
known metric's components. Nevertheless, further, we will fix a gauge for the gravitational action adding some $L_{GF}$ fixing term to the $L_{EH}$ gravitational Lagrangian.

\section{Effective action for scalar particle in a gravitational field}\label{GravPart}

 A high energy scattering of two particles is a non-local in rapidity space process, see \cite{LipatovEff,LipatovEff1,LipatovEff2,Our,Our01} for the descriptions of the QCD high energy scattering processes in the framework of effective action.
In general we have to distinguish three different rapidity regions of the scattering. The  two are at the rapidity edges, at $y\,=\,0$ and $y\,=\,Y$ rapidities, the Lagrangians we have there are separately local in rapidity space and define impact-factors of the scattering process. The third region is in any rapidity cluster in between the edges, it's Lagrangian determines a dynamic of the t-channel Reggeons. The fields we have are the Reggeon scalar and gravity fields, on-shell solutions of free equations of motion (i.e. classical "in" and "out" fields of both kinds) and of-shell fluctuations which we need if we consider quantum loops corrections to any kind of amplitude.

 Let's now consider a rapidity interval, small in comparison to the whole rapidity of the process, where we have two scalar field $\phi$
and $\tph$ move in opposite light-cone directions in a c.m. frame.   
Correspondingly there are two types of the off-shell Reggeon scalar fields we have which we denote as $\mA$ and $\tmA$. The generalization of the \eq{SW1} Lagrangian 
we obtain by the replace 
\beqar\label{qsp1}
\phi\,&\rightarrow&\,\phi\,+\,\tph\,;
\\
\mA\,&\rightarrow&\,\mA\,+\,\tmA\,;\,\,\,\,\D_{+}\mA\,=\,0\,,\,\,\,\,\D_{-}\tmA\,=\,0\,
\label{qsp2}
\eeqar
there.
We have therefore
\beqar\label{qsp3}
L/\sqrt{-g}\,&=&\,L(\phi,\tph)\,+\,L_{int}(\phi,\tph,\mA,\tmA)\,+\,L_{ind}(\mA,\tmA)\,=\,
\\
&=&\,\frac{1}{2}\Le g^{\mu \nu}\D_{\mu}\Le \phi\,+\,\tph \Ra\,\D_{\nu}\Le \phi\,+\,\tph \Ra\,-\,m^2\Le \phi\,+\,\tph \Ra^2\Ra\,-\,
g^{\mu \nu}\D_{\mu}\Le \phi\,+\,\tph \Ra \,\D_{\nu}\Le\mA\,+\,\tmA\Ra\,+\,
\nonumber \\
&+&\,
m^2\phi\Le \phi\,+\,\tph \Ra\Le\mA\,+\,\tmA\Ra\,+\,
g^{\mu \nu}\,\D_{\mu}\Le\mA\,+\,\tmA\Ra\,\D_{\nu}\Le\mA\,+\,\tmA\Ra\,-\,m^2\Le\mA\,+\,\tmA\Ra^2\,-\,
\nonumber\\
&-&\,
\frac{1}{2}\,\Le \eta^{\mu \nu}\,\D_{\mu}\mA\,\D_{\nu}\mA\,-\,m^2\mA^2\Ra\,-\,
\frac{1}{2}\,\Le \eta^{\mu \nu}\,\D_{\mu}\tmA\, \D_{\nu}\tmA\,-\,m^2\tmA^2\Ra\,.
\eeqar
The last two terms in the expression are attended because of two simple reasons. First of all, we do not allow in the Lagrangian a dynamics of Reggeons local in rapidity space so these terms cancel the separated dynamics of $\mA$ and $\tmA$ fields. Secondly,  as we will see further, we have to fix the correct form of the impact factor of the scattering amplitude, it is correct with these both terms included.
The classical LO solutions of equations of motion we obtain are
\beq\label{qsp4}
\phi\,=\,\mA\,,\,\,\,\,\tph\,=\,\tmA\,
\eeq
and the whole \eq{qsp3} Lagrangian reduces to
\beq\label{qsp5}
L/\sqrt{-g}\,=\,
-\frac{1}{2}\,h^{\mu \nu}\,\D_{\mu}\mA\,\D_{\nu}\mA\,
-\frac{1}{2}\,h^{\mu \nu}\,\D_{\mu}\tmA\, \D_{\nu}\tmA\,+\,
g^{\mu \nu}\,\D_{\mu}\mA\,\D_{\nu}\tmA\,-\,m^2\mA\tmA\,.
\eeq
Now we have to account an appearance of the t-channel gravitational Reggeon through the t-channel of-shell scalar Reggeons. We define it similarly to \eq{CE1001}
expression obtaining for the effective Lagrangian:
\beq\label{qsp6}
L_{eff}\,=\,L_{EH}(\eta,h)\,+\,
\Le\eta^{\mu \nu}\, \D_{\mu}\mA\,\D_{\nu}\tmA\,-\,\,m^2\mA\tmA\Ra\,
+\,\frac{1}{2\kappa}\,\D_{i}^{2} \mB_{\mu \nu}\,\Le h^{\mu \nu}\,+\,j^{\,\mu \nu}(h)\Ra\,+\,L_{ind}(B)\,.
\eeq
As before, we have here the Reggeon structure defined by the following identities:
\beqar\label{qsp7}
&\,& \mB_{--}\,\sim\,\D_{-}\mA\,\D_{-}\mA\,;\,\,\,\,B_{-i}\,\sim\,\D_{-}\mA\,\D_{i}\mA\,;\,\,\,\,\D_{+}\mB_{--}\,=\,\D_{+}\mB_{-i}\,=\,0\,;
\\
&\,& \mB_{++}\,\sim\,\D_{+}\tmA\,\D_{+}\tmA\,;\,\,\,\,B_{-i}\,\sim\,\D_{+}\tmA\,\D_{i}\tmA\,;\,\,\,\,\D_{-}\mB_{++}\,=\,\D_{-}\mB_{+i}\,=\,0\,;
\label{qsp8} \\
&\,& \D_{-}\mA\,\D_{+}\tmA\,\Le h^{- +}\,+\,j^{\,- +}(h)\Ra\,=\,0\,\label{qsp801}
\eeqar
and
\beq\label{qsp9}
L_{ind}(B)\,=\,-\,\frac{1}{2\kappa}\,\mB^{\mu \nu}\,\D_{i}^{2} \mB_{\mu \nu}\,.
\eeq
Resolving the classical equations of motion for the $h$ field, we obtain finally the effective Lagrangian which governs a behavior of the t-channel Reggeons:
\beq\label{qsp10}
L_{eff}\,=\,L_{EH}(\eta,\mB)\,+\,
\Le\eta^{\mu \nu}\, \D_{\mu}\mA\, \D_{\nu}\tmA\,-\,\,m^2\mA\tmA\Ra\,
+\,\frac{1}{2\kappa}\,j^{\,\mu \nu}(\mB)\,\D_{i}^{2} \mB_{\mu \nu}\,.
\eeq
The effective current in the expression provides new vertices of interaction of reggeized gravitons absent in the usual Einstein-Hilbert Lagrangian, the vertices in turn provide  new contribution to the propagator of the reggeized gravitons, see further.

 Next we need to account a possibility of the production of scalar particles and gravitons, the produced particles we consider as a solution of free equations of motion and denote as free scalar particles and free gravitons. We preserve the Reggeon form of the reggeized gravity action and 
, see \cite{Faddeev,Faddeev1,Our2}, we describe the production $S$ matrix elements through the following shift of the  fields in the effective action:
\beqar\label{qsp11}
\phi\,&\rightarrow&\,\mA\,+\,\phi_{f}\,;\\
\tph\,&\rightarrow&\,\tmA\,+\,\tph_{f}\,;
\label{qsp12}\\
h_{\mu \nu}\,&\rightarrow&\,\mB_{\mu \nu}\,+\,h_{f\,\mu \nu}\,
\label{qsp13}.
\eeqar
We denote differently the free scalar fields here only in order to classify the possible dependence of the particular free field on different rapidities. 
These both free fields consist correct "in" and "out" states, i.e.
in the effective action we have different types of diagrams in correspondence to the different combinations of the "in" and "out" states.
In the case of the gravitational waves production, the simplest way to introduce them on the stage is firstly to define plane gravitational pp-waves 
similarly to the classical solution defined in Section \ref{CompM} through the following expressions:
\beqar\label{qsp14}
h_{f\,--}\,&=&\,\psi(n_{\mu}x^{\mu},x_{\bot})\,n_{-}\,n_{-}\,,\,\,\,\,n^{\mu}=\,(1,0,0_{\bot})\,,\,\,\,\,\D_{+}\psi\,=\,0\,,\,\,\,\,\D_{i}^{2}\psi\,=\,0\,;
\\
h_{f\,++}\,&=&\,\tilde{\psi}(\tilde{n}_{\mu}x^{\mu},x_{\bot})\,\tilde{n}_{+}\,\tilde{n}_{+}\,,\,\,\,\,\tilde{n}^{\mu}=\,(0,1,0_{\bot})\,,\,\,\,\,\D_{-}\tilde{\psi}\,=\,0\,,\,\,\,\,\D_{i}^{2}\tilde{\psi}\,=\,0\,
\label{qsp15}
\eeqar
and shift correspondingly the gravitational fields:
\beq\label{qsp16}
h_{--}\,\rightarrow\,h_{cl\,--}\,+\,h_{f\,--}\,,\,\,\,\,h_{++}\,\rightarrow\,h_{cl\,++}\,+\,h_{f\,++}\,
\eeq
in the effective Lagrangian. 
Taking into account the corresponding shifts in the induced Lagrangian \eq{qsp9}, we have then for the whole effective Lagrangian:
\beqar\label{qsp19}
L_{eff}\,& = &\,L_{EH}(\eta,\mB)\,+\,L_{0}(\mA,\tmA)\,+\,
\frac{1}{2\kappa}\,j^{\,\mu \nu}(\mB,h_{f})\,\D_{i}^{2} \mB_{\mu \nu}\,-\,
\\
&-&\,
\frac{1}{2}\,\Le \mB_{++}\,+\,h_{f\,++}\Ra\,\Le 2\,\D_{-}\phi_{f}\,\D_{-}\mA\,+\,\D_{-}\phi_{f}\,\D_{-}\phi_{f}\Ra\,-\,
\frac{1}{2}\,\Le \mB_{--}\,+\,h_{f\,--}\Ra\,\Le 2\,\D_{+}\tph_{f}\,\D_{+}\tmA\,+\,\D_{+}\tph_{f}\,\D_{+}\tph_{f}\Ra\,-\,
\nonumber \\
&-&\,
\frac{1}{2}\,j_{\,++}(\mB,h_{f})\,\Le 2\,\D_{-}\phi_{f}\,\D_{-}\mA\,+\,\D_{-}\phi_{f}\,\D_{-}\phi_{f} \Ra\,-\,
\frac{1}{2}\,j_{\,--}(\mB,h_{f})\,\Le 2\,\D_{+}\tph_{f}\,\D_{+}\tmA\,+\,\D_{+}\tph_{f}\,\D_{+}\tph_{f} \Ra\,
\nonumber,
\eeqar
here we accounted that $\D_{i}^{2}h_{f}\,=\,0\,$.
For the construction of the scattering amplitudes we need an impact factor of the process. We can obtain it directly from the \eq{qsp19} Lagrangian
taking there the $\tph\,=\,\tmA\,=\,0$ fields value.  So we have for the Lagrangian (impact factor) at $y\,=\,0\,$ rapidity: 
\beqar\label{qsp20}
L_{eff}(y\,=\,0)\,& = &\,L_{EH}(\eta,\mB,y\,=\,0)\,+\,
\\
&+&\,
\frac{1}{2\kappa}\,j_{--}(\mB,h_{f})\,\D_{i}^{2} \mB_{++}\,-\,
\frac{1}{2}\,\Le \mB_{++}\,+\,h_{f\,++}\,+\,j_{\,++}(\mB,h_{f})\Ra\,\Le 2\,\D_{-}\phi_{f}\,\D_{-}\mA\,+\,\D_{-}\phi_{f}\,\D_{-}\phi_{f}\Ra\,\nonumber.
\eeqar
The Reggeon field $\mB_{--}$ does not present in the expression because of  the $\mB_{--}(y\,=\,0)\,=\,0\,$ boundary condition, see
\cite{Our,Our01,Our1,Our11,Our12} for the details of the RFT construction.
There is a similar expression for the second impact factor at final $y\,=\,Y$ rapidity with corresponding change of the fields performed. 
In general, integrating out all off-shell fields in the partition function, we will obtain an effective action
which generate all possible scattering amplitudes, i.e. $S$ matrix elements, by construction. We discuss that issue in the next Section.

\section{Quantum structure of the effective action}\label{EffA}

 We remind that the amplitudes we calculate are non-local in rapidity space. The calculation procedure we follow is the following 
therefore\footnote{It is similar to calculation scheme introduced in the \cite{Balitsky,Balitsky1}}. First of all, we calculate the effective action 
(impact factor) using the effective action at the rapidity edge:
\beq\label{effa1}
Z_{1}[h_{f},\phi_{f}\,\tph_{f},\mA\,\tmA,\mB]\,=\,e^{\imath\,\Gamma_{1}[h_{f},\phi_{f}\,\tph_{f},\mA\,\tmA,\mB]}\,=\,\int\,
\mathcal{D}\epsilon\,\mathcal{D}\epsilon_{\mu \nu}\,e^{\imath\,S_{1\,eff}}\,,\,\,\,\,S_{1\,eff}\,=\,\int\,d^4 x\,L_{1\,eff}\,,
\eeq
with $\epsilon$ and $\epsilon_{\mu \nu}$ as off-shell fluctuations around classical values of scalar and gravitational fields correspondingly, similarly
the calculation of the second impact factor $\Gamma_{2\,eff}$, at the second edge of the rapidity interval can be performed.
For the in-between rapidity cluster we also can determine the following effective action:
\beq\label{effa2}
Z[h_{f},\phi_{f}\,\tph_{f},\mA\,\tmA,\mB]\,=\,e^{\imath\,\Gamma[h_{f},\phi_{f}\,\tph_{f},\mA\,\tmA,\mB]}\,=\,\int\,
\mathcal{D}\epsilon\,\mathcal{D}\epsilon_{\mu \nu}\,e^{\imath\,S_{eff}}\,,\,\,\,\,S_{1\,eff}\,=\,\int\,d^4 x\,L_{eff}\,,
\eeq
using the \eq{qsp19} Lagrangian. 
Now we can define a general structure of the effective theory. There is the following partition faction of the problem we obtain:
\beq\label{effa3}
Z[h_{f},\phi_{f}\,\tph_{f}]\,=\,e^{\imath\,\Gamma[h_{f},\phi_{f}\,\tph_{f}]}\,=\,\int\,\mathcal{D}\mA\,\mathcal{D}\tmA\,\mathcal{D}\mB_{\mu \nu}\,
e^{\imath\,\Le\Gamma_{1\,eff}^{y=0}\,+\,\Gamma_{2\,eff}^{y=Y}\,+\,\Gamma_{eff}^{y}\Ra}\,,
\eeq
We denote here by rapidity sign the effective action terms because of the following reasons.  First of all, we need to separate the similar vertices which belong to the different action's terms and, secondly,
we underline the structure of the result after the integration over the Reggeon fields, any term arrive must connect the sources with different rapidities. Taking, for example, the bare (no quantum corrections) effective Lagrangians, we have:
\beq\label{effa4}
Z[h_{f},\phi_{f}\,\tph_{f}]\,=\,e^{\imath\,\Gamma[h_{f},\phi_{f}\,\tph_{f}]}\,=\,\int\,\mathcal{D}\mA\,\mathcal{D}\tmA\,\mathcal{D}\mB_{\mu \nu}\,
e^{\imath\,\Le S_{1\,eff}^{y=0}\,+\,S_{2\,eff}^{y=Y}\,+\,S_{eff}^{y}\Ra}\,
\eeq
with
\beqar\label{effa5}
&\,&\,S\,=\,S_{1\,eff}^{y=0}\,+\,S_{2\,eff}^{y=Y}\,+\,S_{eff}^{y}\,=\,
\\
&=& \,-\frac{1}{2}\,
\int\,d^{4} x\,\mB_{++}\,\Le 2\,\D_{-}\phi_{f}\,\D_{-}\mA\,+\,\D_{-}\phi_{f}\,\D_{-}\phi_{f}\Ra_{y=0}\,-\,
\frac{1}{2}\,\int\,d^{4} x\,\mB_{--}\,\Le 2\,\D_{+}\tph_{f}\,\D_{+}\tmA\,+\,\D_{+}\tph_{f}\,\D_{+}\tph_{f}\Ra_{y=Y}\,+\,
\nonumber \\
&+&
\int\,d^{4}z\Le
L_{EH}(\eta,\mB)+L_{0}(\mA,\tmA)-
\frac{1}{2}\,\mB_{++}\Le 2\D_{-}\phi_{f}\,\D_{-}\mA+\D_{-}\phi_{f}\,\D_{-}\phi_{f}\Ra-
\frac{1}{2}\,\mB_{--}\Le 2\D_{+}\tph_{f}\,\D_{+}\tmA+\D_{+}\tph_{f}\,\D_{+}\tph_{f}\Ra\Ra_{y}\,+
\nonumber\\
&+&\,
S_{pr}(h_{f})\,,
\eeqar
where $S_{pr}(h_{f})$ is a LO part of the action related to the LO production of the free gravitons, see \eq{qsp20} expression.
At the first step we obtain an effective action arises after integration over the graviton fields:
\beqar\label{effa6}
&\,&\,S\,=\,\frac{1}{4}\,\int\,d^4 x\,d^4 z\Le 2\,\D_{-}\phi_{f}\,\D_{-}\mA\,+\,\D_{-}\phi_{f}\,\D_{-}\phi_{f} \Ra_{x}\,
<\mB_{++}(x)\,\mB_{--}(z)>_{\Delta y=Y}\,\Le 2\,\D_{+}\tph_{f}\,\D_{+}\tmA\,+\,\D_{+}\tph_{f}\,\D_{+}\tph_{f} \Ra_{z}\,+\,
\\
&+&\,
\frac{1}{4}\,\int\,d^4 x\,d^4 z\Le 2\,\D_{-}\phi_{f}\,\D_{-}\mA\,+\,\D_{-}\phi_{f}\,\D_{-}\phi_{f} \Ra_{x}\,
<\mB_{++}(x)\,\mB_{--}(z)>_{\Delta y=y}\,\Le 2\,\D_{+}\tph_{f}\,\D_{+}\tmA\,+\,\D_{+}\tph_{f}\,\D_{+}\tph_{f} \Ra_{z}\,+\,
\nonumber \\
&+&\,
\frac{1}{4}\,\int\,d^4 x\,d^4 z\Le 2\,\D_{-}\phi_{f}\,\D_{-}\mA\,+\,\D_{-}\phi_{f}\,\D_{-}\phi_{f} \Ra_{x}\,
<\mB_{++}(x)\,\mB_{--}(z)>_{\Delta y=Y-y}\,\Le 2\,\D_{+}\tph_{f}\,\D_{+}\tmA\,+\,\D_{+}\tph_{f}\,\D_{+}\tph_{f} \Ra_{z}\,\nonumber \\
&+&\,
S_{pr}(h_{f})\,,
\eeqar
Here as $<\mB_{++}(x)\,\mB_{--}(y)>$ we denote the graviton propagator which form is defined by $L_{EH}(\eta,\mB)$ Lagrangian.  
If we consider a quasielastic scattering and do not account production of particles inside the rapidity interaval, 
the only term we need to consider in the answer is the first term of \eq{effa6} effective action expression plus
free graviton production part of the action.
So, at the next stage we integrate out the t-channel scalar fields obtaining to LO precsion:
\beqar\label{effa7}
&\,&\,S\,=\,\frac{1}{4}\,
\int d^4 x\,d^4 z\,\,\,\D_{-}\phi_{f}(x)\,\D_{-}\phi_{f}(x)\,<\mB_{++}(x)\,\mB_{--}(z)>\,\D_{+}\tph_{f}(z)\,\D_{+}\tph_{f}(z)\,+\,
\nonumber\\
&+&\,
\int d^4 x\,d^4 z\,\,\,\D_{-}^{2}\phi_{f}(x)\,<\mB_{++}(x)\,\mB_{--}(z)>\,<\mA(x)\,\tmA(z)>\,\D_{+}^{2}\tph_{f}(z)\,+\,
\nonumber\\
&+&\,
\frac{1}{2}\,\int d^4 x\,\,\,\D_{-}\phi_{f}(x)\,\D_{-}\phi_{f}(x)\,h_{f\,++}(x)\,+
\frac{1}{2}\,\int d^4 x\,\,\,\D_{+}\tph_{f}(x)\,\D_{+}\tph_{f}(x)\,h_{f\,--}(x)\,+\,
\nonumber\\
&+&\,
\int d^4 x\,d^4 z\,\,\,\D_{-}^{2}\phi_{f}(x)\,h_{f\,++}(x)\,<\mA(x)\,\tmA(z)>\,h_{f\,--}\,\D_{+}^{2}\tph_{f}(z)\,
\eeqar
here is assumed $<\mA>\,=\,<\tmA>\,=\,0\,$ and the answer is particularly simple of course. This "bare" effective action, therefore, provides the amplitudes of the "bare" high-energy scattering of two scalar particles.
As we see, the general expression for the effective action we obtain in the approach has the following form:
\beq\label{effa8}
\Gamma\,=\,\sum_{m,n\,=\,1}\,\phi_{f}(x_{1})\cdots\phi_{f}(x_{m})\,\mathcal{C}(x_{1}\cdots x_{m},\,z_{1}\cdots z_{m},\,h_{f})\,\tph_{f}(z_{1})\cdots\tph_{f}(z_{n})\,,
\eeq  
here for the each coordinate an separated integration is assumed and the coordinate dependence includes the dependence on the rapidity variable as well.
The vertices of the effective action, therefore, are the amplitudes of the scattering of the scalar particles we are looking for, from the construction it is clear that the amplitudes are non-local in rapidity space.

\section{Impact factor of interaction of scalar particle and Reggeon fields}\label{ImpFac}

 In this Section we discuss the impact factors of the problem. For our further purposes it is enough to calculate only leading (with respect to the gravitational field)  impact factor of the problem. Therefore, first of all, we take $\gamma\,=\,0$ in \eq{CE2003} in the effective currents and do not write in the Section 
expressions for the next order terms of the impact factor, part of them is present in the Appendix \ref{AppB}.

 In general, the impact factor is defined through effective action at the rapidity edge and can include 
one-loop corrections arise from scalar and gravitational fields.
We again begin from the \eq{SW1} adding there the scalar and gravity fields fluctuations. Accounting that $\mB_{--}(y=0)\,=\,0$ at the edge of rapidity, we  obtain for the Lagrangian
at $y\,=\,0$ rapidity, i.e. for the impact factor of the problem:
\beqar\label{effa9}
&\,&L_{eff}(y\,=\,0)\,=\,\frac{1}{2}\,\Le \eta^{\mu \nu}\,\D_{\mu}\epsilon \,\D_{\nu}\epsilon\,-\,m^2\,\epsilon^2 \Ra\,+\,
L_{EH}(\mB,\epsilon_{\mu \nu})\,+\,\frac{1}{2\kappa}\,\Le \epsilon_{--}\,+\,j_{--}\Ra\,\D_{i}^{2} \mB_{++}\,-\,\\
&-&\,
\frac{1}{2}\,\Le \mB_{++}+h_{f\,++}+\epsilon_{++}+j_{\,++}(\mB,h_{f},\epsilon)\Ra\,
\Le 2\D_{-}\phi_{f}\,\D_{-}\mA+\D_{-}\phi_{f}\,\D_{-}\phi_{f}+\D_{-}\epsilon \,\D_{-}\epsilon+
2\D_{-}\epsilon \,\D_{-}\phi_{f}+2\D_{-}\epsilon\,\D_{-}\mA\Ra\,-\,
\nonumber\,\\
&-&\,
\frac{1}{2}\,\epsilon_{--}\,\Le \D_{+}\epsilon \,\D_{+}\epsilon\,+\,2\,\D_{+}\epsilon \,\D_{+}\phi_{f}\,+\,\D_{+}\phi_{f} \,\D_{+}\phi_{f}\Ra\,+\,
\,\epsilon_{-i}\,\Le \D_{+}\epsilon \,\D_{i}\epsilon\,+\,2\D_{+}\epsilon \,\D_{i}\phi_{f}\,+\,2\D_{i}\epsilon \,\D_{+}\phi_{f}\,+\,\D_{+}\phi_{f} \,\D_{i}\phi_{f}\,+\,
\right.
\nonumber \\
&+&\left.
2\D_{+}\epsilon \,\D_{i}\mA\,+\,2\D_{+}\phi_{f} \,\D_{i}\mA\,\Ra\,+\,
\epsilon_{+i}\,\Le \D_{-}\epsilon \,\D_{i}\epsilon\,+\,2\D_{-}\epsilon \,\D_{i}\phi_{f}\,+\,2\D_{i}\epsilon \,\D_{-}\phi_{f}\,+\,\D_{-}\phi_{f} \,\D_{i}\phi_{f}\,+\,
2\D_{-}\epsilon \,\D_{i}\mA\,+\,2\D_{i}\epsilon \,\D_{-}\mA\,+\,
\right. \nonumber \\
&+&\left.
2\D_{-}\phi_{f} \,\D_{i}\mA\,+\,2\D_{i}\phi_{f} \,\D_{-}\mA\,+\,\D_{-}\mA \,\D_{i}\mA\,\Ra\,-\,
\frac{1}{2}\,\epsilon_{i j}\,\Le  
\D_{i}\epsilon \,\D_{j}\epsilon\,+\,2\,\D_{i}\epsilon \,\D_{j}\phi_{f}\,+\,2\,\D_{i}\epsilon \,\D_{j}\mA\,+\,2\,\D_{i}\phi_{f} \,\D_{j}\mA\,+\,
\right.
\nonumber \\
&+&\,
\left.
\,\D_{i}\phi_{f} \,\D_{j}\phi_{f}\,+\,\D_{i}\mA \,\D_{j}\mA\Ra\,\nonumber.
\eeqar
In the presented framework this long expression can be simplified of course. We put attention that in our perturbative scheme we take
\beq\label{effa901}
\D_{i}\mA \,\D_{j}\mA\,=\,\D_{i}\mA \,\D_{-}\mA\,=\,0\,,
\eeq
these fields are sub-leading Reggeons and we can do not account them. Secondly, in paradigm of the Reggeon calculus, we consider the only leading contributions to the scattering amplitudes, i.e. 
the amplitudes with gravitational Reggeon fields attached. Therefore, instead full \eq{effa9}, we use the following expression:
\beqar\label{effa902}
&\,&L_{eff}(y\,=\,0)\,=\,\frac{1}{2}\,\Le \eta^{\mu \nu}\,\D_{\mu}\epsilon \,\D_{\nu}\epsilon\,-\,m^2\,\epsilon^2 \Ra\,+\,
L_{EH}(\mB,\epsilon_{\mu \nu})\,+\,\frac{1}{2\kappa}\,\Le \epsilon_{--}\,+\,j_{--}\Ra\,\D_{i}^{2} \mB_{++}\,-\,\\
&-&\,
\frac{1}{2}\,\Le \mB_{++}+h_{f\,++}+\epsilon_{++}+j_{\,++}(\mB,h_{f},\epsilon)\Ra\,
\Le 2\D_{-}\phi_{f}\,\D_{-}\mA+\D_{-}\phi_{f}\,\D_{-}\phi_{f}+\D_{-}\epsilon \,\D_{-}\epsilon+
2\D_{-}\epsilon \,\D_{-}\phi_{f}+2\D_{-}\epsilon\,\D_{-}\mA\Ra\,-\,
\nonumber\,\\
&-&\,
\frac{1}{2}\,\epsilon_{--}\,\Le \D_{+}\epsilon \,\D_{+}\epsilon\,+\,2\,\D_{+}\epsilon \,\D_{+}\phi_{f}\,+\,\D_{+}\phi_{f} \,\D_{+}\phi_{f}\Ra\,\nonumber.
\eeqar
The last term in the expression we preserved because of the gravity action structure. We have for the $S_{EH}$:
\beq\label{effa10}
S_{EH}\,=\,\frac{1}{4\kappa}\,\int\, d^4 x\,h_{\mu \nu}\,I^{\mu \nu \rho \sigma}\,\Box h_{\rho \sigma}\,,\,\,\,\,
I^{\mu \nu \rho \sigma}\,=\,\frac{1}{2}\Le \eta^{\mu \rho}\eta^{\nu \sigma}\,+\,\eta^{\mu \sigma}\eta^{\nu \rho}\,-\,\eta^{\mu \nu}\eta^{\rho \sigma} \Ra\,
\eeq
obtained with the help of the
\beq\label{effa11}
L_{GF}\,=\,\frac{1}{2\kappa}\,\D_{\mu}\Le h^{\mu \nu}\,-\,\frac{1}{2}\,\eta^{\mu \nu}\,h\Ra\,\D^{\rho}\,\Le h_{\nu \rho}\,-\,\frac{1}{2}\,\eta_{\nu \rho}\,h\Ra\,
\eeq
gauge fixing Lagrangian, see \cite{Magg,GravQF} for example. Preserving the only $\mB_{--}$ and $\mB_{++}$ non-zero classical gravitation weak fields in the framework 
we obtain:
\beqar\label{effa12}
S_{EH}\,& = &\,\frac{1}{2\kappa}\,\int\, d^4 x\,\Le \mB_{--}\,+\,\epsilon_{--}\Ra\,\Box\,\Le \mB_{++}\,+\,\epsilon_{++}\Ra\,-\,
\frac{1}{\kappa}\,\int\, d^4 x\,\epsilon_{+i}\,\Box\,\epsilon_{-i}\,-\,
\frac{1}{2\kappa}\,\int\, d^4 x\,\Le\eta^{+-}\epsilon_{+-}\Ra\,\Box\,\Le \eta^{i j}\epsilon_{i j}\Ra\,+\,
\nonumber \\
&+&\,\frac{I^{ijkl}}{4\kappa}\,\int\, d^4 x\,\epsilon_{ij}\,\Box\,\epsilon_{kl}\,,
\eeqar
here at $y\,=\,0$ we have to apply $h_{f\,--}(y=0)\,=\,\mB_{--}(y=0)\,=\,0$ as well.
The last two terms in \eq{effa12} do not contribute into the partition function of interest and can be factorized out  by the redefinition of the normalization constant of the partition function.
Next we integrate out the scalar field fluctuations in the partition function. The propagator of the scalar field is defined through the following equation:
\beq\label{effa13}
-\,\Le \Box\,+\,m^2\,-\,V_{++}\,\D_{-}^{2}\,-\,\epsilon_{--}\,\D_{+}^{2}\Ra_{x}\,G(x,y)\,=\,\delta(x-y)\,,
\eeq
with
\beqar\label{effa14}
&\,& V_{++}\,=\,\mB_{++}+h_{f\,++}+\epsilon_{++}+j_{\,++}(\mB,h_{f},\epsilon)\,;\\
&\,& j_{\,++}(\mB,h_{f},\epsilon)\,=\,\Le  -\frac{1}{2}\D_{i}\D_{+}^{-1}\Le \mB_{++}+h_{f\,++}+\epsilon_{++}\Ra+\epsilon_{+i}\Ra\,
\Le -\frac{1}{2}\D_{i}\D_{+}^{-1}\Le \mB_{++}+h_{f\,++}+\epsilon_{++}\Ra+\epsilon_{+i} \Ra\,\label{effa1401}
\eeqar
where at $y\,=\,0$ correspondingly we have
\beq\label{effa14001}
j_{\,--}(y\,=\,0)\,=\,\Le  -\frac{1}{2}\D_{i}\D_{-}^{-1}\epsilon_{--}\,+\,\epsilon_{-i}\Ra\,
\Le -\frac{1}{2}\D_{i}\D_{-}^{-1}\epsilon_{--}\,+\,\epsilon_{-i} \Ra\,.
\eeq
Introducing a bare Feynman propagator
\beq\label{effa15}
-\,\Le \Box\,+\,m^2\Ra\,G_{0}(x,y)\,=\,\delta(x-y)\,,\,\,\,G_{0}(x,y)\,=\,\int\,\frac{d^4p}{(2\pi)^{4}}\,\frac{e^{-\imath\,p\,\Le x-y\Ra}}{p^{2}\,-\,m^2}\,,
\eeq
the full propagator can be written in the following form useful for the perturbative calculations:
\beq\label{effa16}
G(x,y)\,=\,G_{0}(x,y)\,-\,\int\,d^{4} z\,G_{0}(x,z)\,\Le \,V_{++}\,\D_{-}^{2}\,+\,\epsilon_{--}\,\D_{+}^{2} \Ra_{z}\,G(z,y).
\eeq
Performing an integration over the scalar field fluctuations we obtain:
\beqar\label{effa17}
&\,&S_{eff}(y\,=\,0)=\int d^{4} x\Le
\frac{1}{2\kappa}\,\epsilon_{--}\,\Box\,\epsilon_{++}-
\frac{1}{\kappa}\,\epsilon_{+i}\,\Box\,\epsilon_{-i}
+\frac{1}{2\kappa}\,j_{--}\,\D_{i}^{2} \mB_{++}-
\right.\\
&-&\,\left.
\frac{1}{2}\,\int\,d^{4}y\,\Le V_{++}\,\Le \D_{-}^{2}\phi_{f}\,+\,\D_{-}^{2}\mA\Ra\,+\,\epsilon_{--}\,\D_{+}^{2}\phi_{f}\Ra_{x}\,G(x,y)\,
\Le V_{++}\,\Le \D_{-}^{2}\phi_{f}\,+\,\D_{-}^{2}\mA\Ra\,+\,\epsilon_{--}\,\D_{+}^{2}\phi_{f}\Ra_{y}
\,-\,
\right.\nonumber \\
&-&\,\left.
\frac{1}{2}\,\epsilon_{--}\,\D_{+}\phi_{f} \,\D_{+}\phi_{f}\,-\,
\frac{1}{2}\,V_{++}\Le 2\D_{-}\phi_{f}\,\D_{-}\mA+\D_{-}\phi_{f}\,\D_{-}\phi_{f}\Ra\Ra\,+\,
\frac{\imath}{2}\,Sp\,\ln\Le 1\,+\,G_{0}\Le V_{++}\,\D_{-}^{2}\,+\,\epsilon_{--}\,\D_{+}^{2} \Ra  \Ra\,\nonumber .
\eeqar
At the next stage we isolate the terms in the effective Lagrangian in respect to the order of the gravitaional $\epsilon_{\mu \nu}$ field.
To the first order, taking $\epsilon_{\mu \nu}\,=\,0$, we have:
\beqar
&\,&S_{eff\,0}(y\,=\,0)=\int d^{4} x\Le\,-\,
\frac{1}{2}\,\int\,d^{4}y\,V_{++\,0}\,\Le \D_{-}^{2}\phi_{f}\,+\,\D_{-}^{2}\mA\Ra\,G(x,y,\epsilon=0)\,
V_{++\,0}\,\Le \D_{-}^{2}\phi_{f}\,+\,\D_{-}^{2}\mA\Ra\,
\,-\,
\right.\nonumber \\
&-&\,\left.
\frac{1}{2}\,V_{++\,0}\Le 2\D_{-}\phi_{f}\,\D_{-}\mA+\D_{-}\phi_{f}\,\D_{-}\phi_{f}\Ra\Ra\,+\,
\frac{\imath}{2}\,Sp\,\ln\Le 1\,+\,G_{0}\, V_{++\,0}\,\D_{-}^{2}\Ra\,\label{effa1701}.
\eeqar
with
\beq\label{effa1702}
V_{++\,0}\,=\,\mB_{++}+h_{f\,++}+\frac{1}{4}\Le \D_{i}\D_{+}^{-1}\Le \mB_{++}+h_{f\,++}\Ra\Ra^{2}\,.
\eeq
The action at rapidity edge above determines the leading order impact factor of interaction of Reggeon fields with the free scalar field.
An expansion of the action with respect to the $h_{f}$ field provides in turn the impact factor of more complex structure, i.e. type of production (diffraction)
impact factor. In the simplest variant of the impact factor we need further, we take this field equal to zero and preserves in the expression only bare linear with respect to reggeized fields terms.
Impact factor calculated to the next orders with respect to the gravitational filed are presented in Appendix \ref{AppB} as mentioned above.

\section{$t$-channel quasi-elastic scattering amplitude}\label{ScatP}

 In this Section we consider the amplitude of $t$-channel reggeized fields at some rapidity cluster in between edges of rapidity interval. In this case the both Reggeon fields are present in the general Lagrangian given by \eq{qsp19}.
For the present paper aims it is enough to consider a quasi-elastic scattering amplitude without production of gravitons and scalar particles, i.e. we take
$h_{f}\,=\,\phi_{f}\,=\,0$ returning to the \eq{qsp10} Lagrangian:
\beq\label{Sca1}
L_{eff}\,=\,L_{EH}(\eta,\mB)\,+\,
\Le\eta^{\mu \nu}\, \D_{\mu}\mA\, \D_{\nu}\tmA\,-\,\,m^2\mA\tmA\Ra\,
+\,\frac{1}{2\kappa}\,j^{\,\mu \nu}(\mB)\,\D_{i}^{2} \mB_{\mu \nu}\,.
\eeq
The scalar Reggeon field propagator is trivial one here, see \eq{effa15}, whereas the form of the propagator of reggeized graviton is not defined yet. Following to 
ideology of \cite{LipatovEff,LipatovEff1,LipatovEff2,Our,Our01,LipatovGrav1}, we preserve in the Lagrangian the only Lipatov's type of the effective vertices (i.e. the singular ones which have the $\D^{-1}$ behavior) and 
obtain for the leading components of the $\mB$ field:
\beqar
S_{eff}(\mB)\,& = & \,
\frac{1}{2\kappa}\,\int\, d^4 x\,\Le \mB_{--}\,+\,\epsilon_{--}\Ra\,\Box\,\Le \mB_{++}\,+\,\epsilon_{++}\Ra\,-\,
\frac{1}{\kappa}\,\int\, d^4 x\,\epsilon_{+i}\,\Box\,\epsilon_{-i}\,+\,
\nonumber \\
&+&\,
\frac{1}{2\kappa}\,\int\, d^4 x\, \mB_{--}\,\,\D_{i}^{2}\,\epsilon_{++}\,+\,
\frac{1}{2\kappa}\,\int\, d^4 x\,\epsilon_{--}\,\D_{i}^{2}\,\mB_{++}\,-\,
\nonumber\\
&-&
\frac{1}{8\kappa}\int\, d^4 x\,\D_{i}^{2} \mB_{++}
\left[4\epsilon_{-i}\D_{i}\D_{-}^{-1}\Le \mB_{--}+\epsilon_{--}\Ra-\Le \D_{i}\D_{-}^{-1}\Le \mB_{--}+\epsilon_{--}\Ra\Ra^{2}\right]\,-\,
\nonumber \\
&-&
\frac{1}{8\kappa}\int\, d^4 x\,\D_{i}^{2} \mB_{--}
\left[4\epsilon_{+i}\D_{i}\D_{+}^{-1}\Le \mB_{++}+\epsilon_{++}\Ra-\Le \D_{i}\D_{+}^{-1}\Le \mB_{++}+\epsilon_{++}\Ra\Ra^{2}\right]\,-\,
\nonumber \\
&-&\,
\frac{\gamma}{2\kappa}\,\int\, d^4 x\,\D_{i}^{2} \mB_{++}\,\Le \epsilon_{--}\,+\,\mB_{--}\Ra\,
\Le\D_{j}\D_{+}^{-1}\Le \epsilon_{+j}-\frac{1}{2}\D_{j}\D_{+}^{-1}\Le \mB_{++}+\epsilon_{++}\Ra\Ra\Ra\,-\,
\nonumber \\
&-&\,
\frac{\gamma}{2\kappa}\,\int\, d^4 x\,\D_{i}^{2} \mB_{--}\,\Le\epsilon_{++}\,+\,\mB_{++}\Ra\,
\Le \D_{j}\D_{-}^{-1}\Le \epsilon_{-j}-\frac{1}{2}\D_{j}\D_{-}^{-1}\Le \mB_{--}+\epsilon_{--}\Ra\Ra\Ra\,
\label{Sca2},
\eeqar
here the effective current is accounted in \eq{CE2003} form
and we remind that $\gamma$ in the expression is only a numerical constant which value we will discuss later.
The Lagrangian determines the effective vertices of the theory. 
In order to explore a possible reggeization of the $\mB$ fields propagator, we need to calculate an one loop self-energy correction to it. 
We have for the graviton reggeized fields then:
\beqar
S_{eff}(\mB)\,& = &\,-\,
\frac{1}{2\kappa}\,\int\, d^4 x\,\mB_{--}\,\D_{i}^{2}\,\mB_{++}\,+\,
\frac{1}{2\kappa}\,\int\, d^4 x\,\epsilon_{--}\,\Box\,\epsilon_{++}\,+\,
\frac{1}{\kappa}\,\int\, d^4 x\,\eta^{ij}\,\epsilon_{+i}\,\Box\,\epsilon_{-j}\,+\,
\nonumber \\
&+&\,
\frac{\gamma}{4\kappa}\,\int\, d^4 x\,\epsilon_{--}\,\Le
\D_{i}^{2}\,\mB_{++}\,\D_{j}^{2}\,\D_{+}^{-1}\,\D_{+}^{-1}\,+\,\D_{j}^{2}\,\D_{-}^{-1}\,\D_{-}^{-1}\,\D_{i}^{2}\,\mB_{--}\,
\Ra\,\epsilon_{++}\,-\,
\nonumber \\
&-&\,
\frac{1}{2\kappa}\,\int\, d^4 x\,\epsilon_{--}\,\Le \D_{i}^{2} \mB_{++}\, \D_{j}\,\D_{-}^{-1}\Ra\,\epsilon_{-j}\,-\,
\frac{\gamma}{2\kappa}\,\int\, d^4 x\,\epsilon_{--}\,\Le \D_{i}^{2} \mB_{++}\, \D_{j}\,\D_{+}^{-1}\Ra\,\epsilon_{+j}\,-\,
\nonumber \\
&-&\,
\frac{1}{2\kappa}\,\int\, d^4 x\,\epsilon_{++}\,\Le \D_{i}^{2} \mB_{--}\, \D_{j}\,\D_{+}^{-1}\Ra\,\epsilon_{+j}\,-\,
\frac{\gamma}{2\kappa}\,\int\, d^4 x\,\epsilon_{++}\,\Le \D_{i}^{2} \mB_{--}\, \D_{j}\,\D_{-}^{-1}\Ra\,\epsilon_{-j}\,+\,
\nonumber \\
&+&\,
\frac{1}{8\kappa}\,\int\, d^4 x\,\D_{i}^{2} \mB_{++}\, \Le \D_{j}\,\D_{-}^{-1}\,\epsilon_{--}\Ra^{2}\,+\,
\frac{1}{8\kappa}\,\int\, d^4 x\,\D_{i}^{2} \mB_{--}\, \Le \D_{j}\,\D_{+}^{-1}\,\epsilon_{++}\Ra^{2}\,-\,
\nonumber \\
&-&\,
\frac{1}{2\kappa}\,\int\, d^4 x\,\epsilon_{-j}\,\Le \D_{i}^{2} \mB_{++}\, \D_{j}\D_{-}^{-1}B_{--}\Ra\,-\,
\frac{1}{2\kappa}\,\int\, d^4 x\,\epsilon_{+j}\,\Le \D_{i}^{2} \mB_{--}\, \D_{j}\D_{+}^{-1}B_{++}\Ra\,-\,
\nonumber \\
&-&
\frac{\gamma}{2\kappa}\,\int\, d^4 x\, \D_{i}^{2} \mB_{--}\,\mB_{++}\, \D_{j}\,\D_{-}^{-1}\epsilon_{-j}\,-\,
\frac{\gamma}{2\kappa}\,\int\, d^4 x\, \D_{i}^{2} \mB_{++}\,\mB_{--}\, \D_{j}\,\D_{+}^{-1}\epsilon_{+j}\,+\,
\nonumber \\
&+&
\frac{1}{4\kappa}\,\int\, d^4 x\, \D_{i}^{2} \mB_{++}\, \D_{j}\D_{-}^{-1}B_{--}\,\D_{j}\D_{-}^{-1}\epsilon_{--}\,+\,
\frac{1}{4\kappa}\,\int\, d^4 x\, \D_{i}^{2} \mB_{--}\, \D_{j}\D_{+}^{-1}B_{++}\,\D_{j}\D_{-}^{-1}\epsilon_{++}\,-\,
\nonumber \\
&+&
\frac{\gamma}{4\kappa}\,\int\, d^4 x\, \D_{i}^{2} \mB_{--}\,\mB_{++}\, \D_{j}^{2}\D_{-}^{-1}\D_{-}^{-1}\epsilon_{--}\,+\,
\frac{\gamma}{4\kappa}\,\int\, d^4 x\, \D_{i}^{2} \mB_{++}\,\mB_{--}\, \D_{j}^{2}\D_{+}^{-1}\D_{-}^{-1}\epsilon_{++}\,+\,
\nonumber \\
&+&
\frac{\gamma}{4\kappa}\,\int\, d^4 x\, \D_{i}^{2} \mB_{++}\,\D_{j}^{2}\D_{+}^{-1}\D_{+}^{-1}\mB_{++}\, \epsilon_{--}\,+\,
\frac{\gamma}{4\kappa}\,\int\, d^4 x\, \D_{i}^{2} \mB_{--}\,\D_{j}^{2}\D_{-}^{-1}\D_{-}^{-1}\mB_{--}\, \epsilon_{++}\,+\,
\nonumber \\
&+&\,
\frac{1}{8\kappa}\,\int\, d^4 x\,\D_{i}^{2} \mB_{++}\,\Le \D_{i}\D_{-}^{-1}\mB_{--}\Ra^{2}\,+\,
\frac{1}{8\kappa}\,\int\, d^4 x\,\D_{i}^{2} \mB_{--}\,\Le \D_{i}\D_{+}^{-1}\mB_{++}\Ra^{2}\,+\,
\nonumber \\
&+&
\frac{\gamma}{4\kappa}\,\int\, d^4 x\, \D_{i}^{2} \mB_{--}\,\mB_{++}\, \D_{j}^{2}\D_{-}^{-1}\D_{-}^{-1}\mB_{--}\,+\,
\frac{\gamma}{4\kappa}\,\int\, d^4 x\, \D_{i}^{2} \mB_{++}\,\mB_{--}\, \D_{j}^{2}\D_{-}^{-1}\D_{-}^{-1}\mB_{++}\,.
\label{Sca3}
\eeqar
From the symmetry of the obtained expression we see, that indeed non-zero value of the constant we can take is $\gamma\,=\,1$, we will consider this value of the constant further.
The simplest way to integrate out the fluctuations it is to put attention that the $\epsilon_{\pm i}$ fluctuations are integrated out trivially. In this Section we introduce
bare propagators defined as in an usual QFT where the $1/2$ coefficient is assumed in front of the kinetic term, we have correspondingly:
\beqar\label{Sca301}
&\,&G_{0\,-i+j}\,=\,-\,\kappa\,\eta_{ij}\,\int\,\frac{d^{4}p}{(2\pi)^{4}}\,\frac{e^{-\imath\,p\,(x-y)}}{p^{2}\,+\,\imath\varepsilon}\,;
\\
&\,&G_{0\,++--}\,=\,-\,2\,\kappa\,\int\,\frac{d^{4}p}{(2\pi)^{4}}\,\frac{e^{-\imath\,p\,(x-y)}}{p^{2}\,+\,\imath\varepsilon}\,\label{Sca30101},
\eeqar
 we obtain after that integration:
\beqar
S_{eff}(\mB)\,& = &\,-\,
\frac{1}{2\kappa}\,\int\, d^4 x\,\mB_{--}\,\D_{i}^{2}\,\mB_{++}\,+\,
\frac{1}{2\kappa}\,\int\, d^4 x\,\epsilon_{--}\,\Box\,\epsilon_{++}\,+\,
\nonumber \\
&+&\,
\frac{1}{4\kappa}\,\int\, d^4 x\,\epsilon_{--}\,\Le
\D_{i}^{2}\,\mB_{++}\,\D_{j}^{2}\,\D_{+}^{-1}\,\D_{+}^{-1}\,+\,\D_{j}^{2}\,\D_{-}^{-1}\,\D_{-}^{-1}\,\D_{i}^{2}\,\mB_{--}\,
\Ra\,\epsilon_{++}\,+\,
\nonumber \\
&+&\,
\frac{1}{8\kappa}\,\int\, d^4 x\,\D_{i}^{2} \mB_{++}\, \Le \D_{j}\,\D_{-}^{-1}\,\epsilon_{--}\Ra^{2}\,+\,
\frac{1}{8\kappa}\,\int\, d^4 x\,\D_{i}^{2} \mB_{--}\, \Le \D_{j}\,\D_{+}^{-1}\,\epsilon_{++}\Ra^{2}\,-\,
\nonumber \\
&-&\,
\frac{1}{2\kappa^2}\,
\int\, d^4 x\,\int d^4 y\,\left[\epsilon_{--}\,\Le \D_{i}^{2} \mB_{++}\, \D_{j}\,\D_{-}^{-1}\Ra\right]_{x}\,
\left[ \epsilon_{++}\,\Le \D_{i}^{2} \mB_{--}\, \D_{k}\,\D_{+}^{-1}\Ra\right]_{y}\,G_{0\,-j+k}(x,y)\,-\,
\nonumber \\
&-&\,
\frac{1}{4\kappa^2}\,
\int\, d^4 x\,\int d^4 y\,\left[\epsilon_{--}\,\Le \D_{i}^{2} \mB_{++}\, \D_{j}\,\D_{-}^{-1}\Ra\right]_{x}\,
\left[ \epsilon_{--}\,\Le \D_{i}^{2} \mB_{++}\, \D_{k}\,\D_{+}^{-1}\Ra\right]_{y}\,G_{0\,-j+k}(x,y)\,-\,
\nonumber \\
&-&\,
\frac{1}{4\kappa^2}\,
\int\, d^4 x\,\int d^4 y\,\left[\epsilon_{++}\,\Le \D_{i}^{2} \mB_{--}\, \D_{j}\,\D_{+}^{-1}\Ra\right]_{x}\,
\left[ \epsilon_{++}\,\Le \D_{i}^{2} \mB_{--}\, \D_{k}\,\D_{-}^{-1}\Ra\right]_{y}\,G_{0\,+j-k}(x,y)\,+\,
\nonumber \\
&+&\,
\,S_{int}(\mB,\epsilon)\,.
\label{Sca302}
\eeqar
Here the $S_{int}(\mB,\epsilon)$ represents terms of interaction of reggeized gravitons before an integration over the $\epsilon_{\pm}$ fluctuations,
it is represented in Appendix ~\ref{AppD}. The terms there are important for the construction of RFT (Regge Field Theory) but we do not need them for the calculations of the 
propagator of reggeized gravitons. The remained terms of \eq{Sca302} in general could contribute to the expression of the propagator, in order to clarify their possible contribution 
we need to determine the full bare $G_{++--}$ propagator.
So, the action we have provides the following equation for the full bare graviton's Green's function:
\beq\label{Sca4}
S^{\mu\nu\,\mu_1\nu_1}\,G_{\,\mu_1\nu_1\,\rho\sigma}\,=\,\delta^{\,\mu\nu}_{\,\rho\sigma}
\eeq
with
\beq\label{Sca5}
S^{\mu\nu\,\mu_1\nu_1}\,=\,\frac{\delta^{2}\,S_{eff}}{\delta \epsilon_{\mu\nu}\,\delta \epsilon_{\mu_1\nu_1}}
\eeq
see \eq{effa22} above.
The system of equations we obtain correspondingly is:
\beqar\label{Sca6}	
&\,& \Le S^{\,--\,++}_{0}\,+\,S^{\,--\,++}_{1}+\,S^{\,--\,++}_{2}\Ra\, G_{\,++\,--}\,+\,S^{\,--\,--}\,G_{\,--\,--}\,=\,1 \,\nonumber \\
&\,& \Le S^{\,--\,++}_{0}\,+\,S^{\,--\,++}_{1}+\,S^{\,--\,++}_{2}\Ra\,G_{\,++\,++}\,+\,S^{\,--\,--}\,G_{\,--\,++}\,=\,0 \,\nonumber \\
&\,& S^{\,++\,++}\,G_{\,++\,--}\,+\,\Le S^{\,++\,--}_{0}\,+\,S^{\,++\,--}_{1}\,+\,S^{\,--\,++}_{2}\Ra\,G_{\,--\,--}\,=\,0 \,\nonumber \\
&\,& S^{\,++\,++}\,G_{\,++\,++}\,+\,\Le S^{\,++\,--}_{0}\,+\,S^{\,++\,--}_{1}\,+\,S^{\,--\,++}_{2}\Ra\,G_{\,--\,++}\,=\,1\,,
\eeqar
the new equations in the system are due the new effective vertices in the action, as $S_{n}$ we denote a type of the effective vertex which depend on $n$ external fields, see Appendix \ref{AppC}.
Further, for the shortness,  we will use a "condensed" notation denoting two similar indexes by one: $(\pm \pm)\,\rightarrow\,(\pm)$. 
Using the bare propagator definition
\beq\label{Sca7}
S^{-+}\,G_{0\,+-}\,=\,\frac{1}{2\kappa}\,\Box\,G_{0\,+-}\,=\,1\,,
\eeq
see \eq{effa22} solution, we have:
\beqar\label{Sca8}	
&\,& G_{\,--}\,=\,-\,G_{\,0\,-+}\,\Le 1\,+\,G_{\,0\,-+}\,\Le S_{1}^{+-}\,+\,S^{+-}_{2}\Ra \Ra^{-1}\,S^{\,++}\,G_{\,+-} \,; \\
&\,& G_{\,++}\,=\,-\,G_{\,0\,+-}\,\Le 1\,+\,G_{\,0\,+-}\,\Le S_{1}^{-+}\,+\,S^{-+}_{2}\Ra \Ra^{-1}\,S^{\,--}\,\,G_{\,-+} \,;\nonumber \\
&\,& G_{+-}\,=\,\Le 1\,+\,G_{\,0\,+-}\,\Le S^{-+}_{1}\,+\,S^{-+}_{2}\Ra-\,G_{\,0\,+-}\,S^{\,--}\,G_{\,0\,-+}\,
\Le 1\,+\,G_{\,0\,-+}\,\Le S_{1}^{+-}\,+\,S^{+ -}_{2}\Ra \Ra^{-1}\,S^{++} \Ra^{-1}\,G_{\,0\,+-} \nonumber\,.
\eeqar
The precise form of the vertices is given in Appendix \ref{AppC} calculations. Now we can interact out the the $\epsilon_{\pm}$ fluctuations obtaining
very simple expression:
\beq\label{Sca901}
S_{eff}(\mB)\, =  \,-\,\frac{1}{2\kappa}\,\int\, d^4 x\,\mB_{--}\,\D_{i}^{2}\,\mB_{++}\,-\,\frac{\imath}{2}\, Sp\,\ln\Le G \Ra\,+\,
S_{int}(\mB)\,,
\eeq 
with expression for $S_{int}(\mB)$ is written in Appendix \ref{AppD}.
The terms there mostly represent different non-local
interactions between the reggeized graviton fields. Whereas they are important for calculations of unitarity corrections in the RFT framework,
we do not need them for the calculation of the propagator of reggeized graviton, all we need are the terms which will not vanish in the limit of the 
zero reggeized fields after the differentiation of the action with respect to $\mB_{++}$ and $\mB_{--}$.

  Namely, the  propagator of the reggeized gravitons $\mB_{+}$ and $\mB_{-}$\footnote{Here we again use the shortened notations for the indexes.} is defined through an effective vertex 
\beq\label{Sca10}
\Le\,S_{x y}\,\Ra^{+\,-}\,=\,S_{x y}\,=\,\Le\,\frac{\delta^{2}\,S_{eff}}{\delta \mB_{+\,x}\,\delta \mB_{-\,y}}\,\Ra_{\mB_{+},\,\mB_{-}\,=\,0}\,.
\eeq
in the \eq{Sca901} effective action.
This vertex has the following form
\beq\label{Sca1001}
S^{+\,-}(x^{+},x_{\bot};y^{-},y_{\bot})\,=\,\int\,dx^{-}\,\int\,dy^{+}\,\tilde{S}^{+-}(x,y)
\eeq
that provides 
\beq\label{Sca14}
\tilde{S}^{+-}\Le x,y \Ra\,=\,\delta(y^{-}\,-\,x^{-})\,\delta(x^{+}\,-\,y^{+})\,S(x_{\bot},y_{\bot})\,.
\eeq
The LO vertex in the formalism we obtain therefore is:
\beq\label{Sca11}
S_{x y\,0}\,=\,-\,\frac{1}{\kappa}\,\delta_{x_{\bot} y_{\bot}}\,\D_{i\,x}^{2}\,.
\eeq
The bare propagator satisfies the following equation in turn:
\beq\label{Sca12}
\int\,d^{4} z\,\Le\,\tilde{S}_{x z\,0}\Ra^{-\,+}\,\Le\,\tilde{D}_{z y\,0}\Ra_{+\,-}\,=\,\delta^{4}_{x y}\,
\eeq
with
\beq\label{Sca15}
\tilde{D}_{0\,+ -}\Le x^{+},\,x^{-},\,x_{\bot}\,;\,y^{+},\,y^{-},\,y_{\bot}  \Ra\,=\,\delta(y^{-}\,-\,x^{-})\,\delta(x^{+}\,-\,y^{+})\,
\,D_{0\,+ -}(x_{\bot},\,y_{\bot})\,.
\eeq
that provides 
\beq\label{Sca13}
D_{0\,+ -}(x_{\bot},\,y_{\bot})\,=\,D_{0}(x_{\bot},\,y_{\bot})\,=\,\kappa\,\int\,\frac{d^{2}p}{(2\pi)^{2}}\,\frac{e^{-\imath\,p_{i}\,(x^{i}-y^{i})}}{p^{2}_{\bot}}\,.  
\eeq
Correspondingly,
the full propagator of reggeized gluons is defined through
\beq\label{Sca16}
\int\,d^{4} z\,\Le\,\tilde{S}_{x z}\Ra^{-\,+}\,\Le\,\tilde{D}_{z y}\Ra_{+\,-}\,=\,\delta^{4}_{x y}\,
\eeq
with
\beq\label{Sca17}
\tilde{D}_{z y}\,=\,\sum_{k=0}\,\tilde{D}_{z y;\,k}
\eeq
as a full perturbative kernel of interacting gravitons.
The solution of the propagator, therefore, can be written in the form of the following perturbative series:
\beq\label{Pro6}
 \tilde{D}_{x y}\,=\,\tilde{D}_{x y\,0}\,-\,\int\,d^{4} z\,\int\,d^{4} w\, \tilde{D}_{0\,x z}\,
\Le\,\sum_{k\,=\,1}\, \tilde{S}_{z w;\,k}\,\Ra\, \tilde{D}_{w y}\,.
\eeq
To the leading order precision we need to calculate the $S_{z w;\,1}$ kernel, the calculations are presented in Appendix \ref{AppB}. 
Accounting that
\beq\label{Sca18}
\tilde{D}_{+ -}\Le x^{+},\,x^{-},\,x_{\bot}\,;\,y^{+},\,y^{-},\,y_{\bot}  \Ra\,=\,\delta(y^{-}\,-\,x^{-})\,\delta(x^{+}\,-\,y^{+})\,
\,D_{+ -}(x_{\bot},\,y_{\bot})\,
\eeq
we obtain to this order:
\beq\label{Sca19}
 D(x_{\bot},y_{\bot} )\,=\,D_{0}(x_{\bot},y_{\bot})\,-\,\int\,d^{2} z_{\bot}\,\int\,d^{2} w_{\bot}\, D_{0}(x_{\bot},z_{\bot})\,
\Le\,S_{11}(z_{\bot},w_{\bot})\,+\,S_{12}(z_{\bot},w_{\bot})\Ra\, D(w_{\bot},y_{\bot})\,.
\eeq
Performing Fourier transform, using results of Appendix ~\ref{AppC} and writing explicitly the $\kappa$ in front of all propagators, we have:
\beq\label{Sca19001}
 D(q_{\bot},\xi)\,=\,\frac{1}{q_{\bot}^{2}}\,-\,\frac{\kappa}{8\pi}\,\int^{\xi}\frac{dp_{-}}{p_{-}}\,
\int\,\frac{d^2 p_{\bot}}{(2\pi)^{2}}\,\frac{q_{\bot}^{2}}{p_{\bot}^{2}\,\Le p_{\bot}\,-\,q_{\bot} \Ra^{2}}\,
\Le  2\,\Le \gamma+1\Ra\,\frac{\Le p_{i}(p-q)_{i}\Ra^{2}}{p_{\bot}^{2}}\,+\,2\,\gamma\,p_{\bot}^{2}\,-\,q_{\bot}^{2}\Ra\, D(q_{\bot},p_{-}).
\eeq
As mentioned, the $\gamma\,=\,0$ case reproduces the \cite{LipatovGrav,LipatovGrav01,LipatovGrav02,LipatovGrav03,LipatovGrav04,LipatovGrav05,LipatovGrav06,LipatovGrav1} result if we also will account the different coupling used in the \cite{LipatovGrav,LipatovGrav01,LipatovGrav02,LipatovGrav03,LipatovGrav04,LipatovGrav05,LipatovGrav06,LipatovGrav1} papers.
Namely, redefining the coupling in the front of the Lagrangian as $\kappa\,\rightarrow\,2\kappa^2$, as it defined in 
\cite{LipatovGrav,LipatovGrav01,LipatovGrav02,LipatovGrav03,LipatovGrav04,LipatovGrav05,LipatovGrav06,LipatovGrav1}, we precisely reproduce the
\cite{LipatovGrav,LipatovGrav01,LipatovGrav02,LipatovGrav03,LipatovGrav04,LipatovGrav05,LipatovGrav06,LipatovGrav1} answer. Now, introducing a rapidity variable
\beq\label{Sca20}
y\,=\,\frac{1}{2}\,\ln(\Lambda p_{-})
\eeq
and redefining  an integration over $p_{-}$ as integral over some rapidity cluster 
we obtain for the \eq{Sca19}
\beq\label{Sca19002}
 D(q_{\bot},\msfsl{y})\,=\,\frac{1}{q_{\bot}^{2}}\,-\,\frac{\kappa}{4\pi}\,\int^{\msfsl{y}}_{0}\,dy\,
\int\,\frac{d^2 p_{\bot}}{(2\pi)^{2}}\,\frac{q_{\bot}^{2}}{p_{\bot}^{2}\,\Le p_{\bot}\,-\,q_{\bot} \Ra^{2}}\,
\Le  2\,\Le \gamma+1\Ra\,\frac{\Le p_{i}(p-q)_{i}\Ra^{2}}{p_{\bot}^{2}}\,+\,2\,\gamma\,p_{\bot}^{2}\,-\,q_{\bot}^{2}\Ra\, D(q_{\bot},y)\,.
\eeq
Introducing a trajectory function of the reggeized graviton
\beq\label{Sca20001}
\omega(q_{\bot}^{2})\,=\,-\,\frac{\kappa}{4\pi}\,
\int\,\frac{d^2 p_{\bot}}{(2\pi)^{2}}\,\frac{q_{\bot}^{2}}{p_{\bot}^{2}\,\Le p_{\bot}\,-\,q_{\bot} \Ra^{2}}\,
\Le  2\,\Le \gamma+1\Ra\,\frac{\Le p_{i}(p-q)_{i}\Ra^{2}}{p_{\bot}^{2}}\,+\,2\,\gamma\,p_{\bot}^{2}\,-\,q_{\bot}^{2}\Ra\,.
\eeq
At the $p_{\bot}\,\ll\,q_{\bot}$ limit the trajectory reduces to 
\beq\label{Sca2001}
\omega(q_{\bot}^{2})\,=\,\kappa\,\frac{q_{\bot}^{2}}{16\pi^3}\,
\int^{q_{\bot}^{2}}\,\frac{d^2 p_{\bot}}{p_{\bot}^{2}}\,\approx\,\kappa\,\frac{q_{\bot}^{2}}{16\pi^2}\,\ln(q_{\bot}^{2}/q_{0\bot}^{2})
\eeq
and it is almost linear with respect to $q_{\bot}^{2}$.
Next, taking an derivative with respect to $\msfsl{y}$ we obtain:
\beq\label{Sca20002}
\frac{\D\,D(q_{\bot},\msfsl{y})}{\D\,\msfsl{y}}\,=\,\omega(q_{\bot}^{2})\,D(q_{\bot},\msfsl{y})\,.
\eeq
In turn, therefore, the propagator of the reggeized graviton for the $Y\,=\,\ln(s/s_{0})\,>\,\msfsl{y}\,>0$ overall rapidity interval of the scattering, acquires the following form:
\beq\label{Sca21}
D(q_{\bot},s)\,=\,\frac{\kappa}{q_{\bot}^{2}}\,\Le \frac{s}{s_{0}}\Ra^{\omega(q_{\bot}^{2})}\,=\,\frac{\kappa}{q_{\bot}^{2}}\,e^{Y\,\omega(q_{\bot}^{2})}\,,
\eeq
which is propagator of reggeized gravitons with trajectory calculated firstly in \cite{LipatovGrav,LipatovGrav01,LipatovGrav02,LipatovGrav03,LipatovGrav04,LipatovGrav05,LipatovGrav06,LipatovGrav1}.

\section{Leading order amplitude of scattering of two scalar particles}\label{LOA} 

 Following to ideas of \cite{Faddeev,Faddeev1}, see also \cite{Our,Our01}, the general amplitude (or $S$-matrix element) of the scattering is represented by \eq{effa8} expression.
For our purposes it is enough to take the leading order answer for scatterong of two scalar particles through the one $t$-channel reggeized graviton exchange:
\beq\label{Ampl1}
M_{2\rightarrow2}\,=\,\frac{1}{4}\,
\int d^4 x\,d^4 z\,\,\,\D_{-}\phi_{1f}(x)\,\D_{-}\phi_{2f}(x)\,<\mB_{++}(x)\,\mB_{--}(z)>\,\D_{+}\tph_{1f}(z)\,\D_{+}\tph_{2f}(z)\,,
\eeq
see \eq{effa7} with
\beq\label{Ampl2}
<\mB_{++}(x)\,\mB_{--}(z)>\,=\,-\,\imath\,D(x,z)
\eeq
aka propagator of reggeized gravitons, see also \eq{Sca18}. The free (classical) fields we consider are on-shell plane waves, we choose:
\beqar\label{Ampl3001}
\phi_{1f}(x)\,&=&\,a_{p}\,e^{-\imath\,p_{-} x^{-}\,+\,\imath\,p_{i} x_{i}};
\\
\phi_{2f}(x)\,&=&\,b_{p_{1}}\,e^{\imath\,p_{1-} x^{-}\,-\,\imath\,p_{1i} x_{i}};
\label{Ampl31} \\
\tph_{1f}(z)\,&=&\,\tilde{a}_{k}\,e^{-\imath\,k_{+} z^{+}\,+\,\imath\,k_{i} z_{i}};
\label{Ampl32} \\
\tph_{2f}(z)\,&=&\,\tilde{b}_{k_{1}}\,e^{\imath\,k_{1+} z^{+}\,-\,\imath\,k_{1i} z_{i}},
\label{Ampl33}
\eeqar
where from the kinematics of the scattering process we have: 
\beq\label{Ampl5}
p_{-}\,=\,p_{1-}\,=\,\sqrt{\frac{s}{2}}\,\Le 1,0,0_{\bot}\Ra\,,\,\,\,\,k_{+}\,=\,k_{1+}\,=\,\sqrt{\frac{s}{2}}\,\Le 0,1,0_{\bot}\Ra\,.
\eeq
Consequently we obtain for the amplitude after it's renormalization on the $a_{p} \tilde{a}_{k} b_{p_{1}} \tilde{b}_{k_{1}}$ numerical factor:
\beq\label{Ampl6}
M_{2\rightarrow2}\,= \,-\,\imath\,\frac{s^{2}}{16}\,(2\pi)^{2}\,\delta(p_{-}\,-\,p_{1-})\,\delta(k_{+}\,-\,k_{1+})\,\int d^{2}x_{\bot}\,
\int d^{2}z_{\bot}\,D(x_{\bot}\,-\,z_{\bot})\,e^{-\,\imath\,p_{1i} x_{i}\,+\,\imath\,p_{i} x_{i}}\,e^{-\imath\,k_{1i} z_{i}\,+\,\imath\,k_{i} z_{i}}\,.
\eeq
Performing the following change of variables:
\beq\label{Ampl7}
p_{\bot}\,-\,p_{1\bot}\,=\,q_{\bot}\,;\,\,\,
k_{1\bot}\,-\,k_{\bot}\,=\,q_{1\bot}\,;\,\,\,
x_{\bot}\,-\,z_{\bot}\,=\,w_{\bot}\,,
\eeq
we obtain finally:
\beq\label{Ampl8}
M_{2\rightarrow2}\,= \,-\,\imath\,\frac{s^{2}}{16}\,(2\pi)^{3}\,\delta(p_{-}\,-\,p_{1-})\,\delta(k_{+}\,-\,k_{1+})\,\delta^{2}(q_{\bot}\,-\,q_{1\bot})\,
D(q_{\bot})\,.
\eeq
It is instructive to compare the result with the Newtonian (low energy) limit of the same t-channel one particle amplitude in scattering of two scalar particles. 
Taking 
$s_{0}\,=\,m^2$ with $m$ as a mass of scalar particle we rewrite \eq{Ampl8} answer as:
\beq\label{Ampl9}
M_{2\rightarrow2}\,= \,-\,\imath\,\kappa\,\frac{m^{4}}{16}\,(2\pi)^{3}\,\delta(p_{-}\,-\,p_{1-})\,\delta(k_{+}\,-\,k_{1+})\,\delta^{2}(q_{\bot}\,-\,q_{1\bot})\,
\frac{1}{q_{\bot}^{2}}\,\Le\frac{s}{m^2}\Ra^{2\,+\,\omega(q_{\bot})}\,.
\eeq
Now, borrowing the answer from the \cite{Donoghue,Donoghue1,Donoghue2,Donoghue3} paper, we have for the ratio of the high energy ($M_{HE}$) and Newtonian ($M_{N}$) amplitudes in the momentum space:
\beq\label{Ampl10}
M_{HE}/M_{N}\,\propto\,\Le \frac{s}{m^2}\Ra^{\omega(q)}\,,
\eeq
here we did not include numerical coefficients for both amplitudes of course. As we see, the reggeization of the propagator change the value of the amplitude, there is an enhancement of the amplitude value for the same values of the masses involved in the scattering. 
Namely, there are some effective values of the masses $m$ in the \eq{Ampl8} expression which we can write through the regular Newtonian particle's masses as 
\beq\label{Ampl11}
m\,\rightarrow\,m\,\Le\frac{s}{m^2}\Ra^{\omega(q_{\bot})/4}\,\gg\,m\,.
\eeq
We see that now the same value of the amplitude as in the Newtonian limit we obtain for the mass which is in 
\beq\label{Ampl12}
\Le\frac{s}{m^2}\Ra^{\omega(q_{\bot})/4}\,
\eeq
times smaller then the mass appear in the regular Newtonian expression. 
Unfortunately, this effect is proportional to $q^{2}$   
and therefore it is negligibly small of course  if we talk about any classical or inclusive processes especially in gravity, see also discussion in the Section \ref{Conc}.

 We note finally, that the \eq{effa8} answer is universal, namely, the classical field can be chosen differently there. We can take the Appendix ~\ref{AppA} solution for example or write 
the classical fields as:
\beqar\label{Ampl13}
\phi_{1f}(x)\,&=&\,\int\,\frac{d^{2}p_{\bot}}{(2\pi)^2}\,a(x^{-},p_{\bot})\,e^{\imath\,p_{i} x_{i}};
\\
\phi_{2f}(x)\,&=&\,\int\,\frac{d^{2}p_{1\bot}}{(2\pi)^2}\,b(x^{-},p_{1\bot})\,e^{-\,\imath\,p_{1i} x_{i}};
\label{Ampl3103} \\
\tph_{1f}(z)\,&=&\,\int\,\frac{d^{2}k_{\bot}}{(2\pi)^2}\,\tilde{a}(z^{+},k_{\bot})\,e^{\imath\,k_{i} z_{i}};
\label{Ampl3203} \\
\tph_{2f}(z)\,&=&\,\int\,\frac{d^{2}k_{1\bot}}{(2\pi)^2}\,\tilde{b}(z^{+},k_{1\bot})\,e^{-\,\imath\,k_{1i} z_{i}}
\label{Ampl3303}
\eeqar
with each from the coefficients of the plane wave expansion for the massive scalar field as a function of the squared C.M. frame energy satisfies Klein-Gordon equation:
\beq\label{Ampl4}
\Le p_{\bot}^{2}\,+\,m^2\Ra\,a_{p}(s,p_{\bot})\,=\,0\,
\eeq
with requested initial and final conditions of the functions assumed which depend on the $x^{-}$ and $x^{+}$ coordinates.
In this case, after the \eq{Ampl7} change of variables, we can introduce impact factors of the scattering defined as following:
\beqar\label{Ampl14}
\Psi(q_{\bot})\,& = &\,\frac{1}{s}\,\int\,\frac{d^{2}p_{\bot}}{(2\pi)^2}\,\int dx^{-}\, \D_{-}a(x^{-},p_{\bot})\,\D_{-}b(x^{-},p_{\bot}\,-\,q_{\bot});
\\
\tilde{\Psi}(q_{\bot})\,& = &\,\frac{1}{s}\,\int\,\frac{d^{2}k_{\bot}}{(2\pi)^2}\,\int dz^{+}\, \D_{+}\tilde{a}(z^{+},k_{\bot})\,
\D_{+} \tilde{b}(z^{+},k_{\bot}\,+\,q_{\bot})\,.
\label{Ampl141}
\eeqar
Taking again $s_{0}\,=\,m^2$ we obtain finally:
\beq\label{Ampl15}
M_{2\rightarrow2}\, =  \,-\,\imath\,\kappa\,\frac{m^{4}}{4}\,
\int\,\frac{d^{2}q_{\bot}}{(2\pi)^2}\,\Psi(q_{\bot})\,\frac{1}{q_{\bot}^{2}}\,\Le\frac{s}{m^2}\Ra^{2\,+\,\omega(q_{\bot})}\,
\tilde{\Psi}(q_{\bot})\,.
\eeq
Yet, the \eq{Ampl10} amplification result is holding in this case as well if we use the same impact factors in each full amplitude expression.

 \section{Conclusion}\label{Conc} 

 In this paper we reproduced an old result of \cite{LipatovGrav,LipatovGrav01,LipatovGrav02,LipatovGrav03,LipatovGrav04,LipatovGrav05,LipatovGrav06} in the framework of Lipatov's effective QFT theory introduced in 
\cite{LipatovGrav1} which describes high energy gravitational scattering in terms of $t$-channel reggeized gravitons. We calculated a trajectory of propagator of the reggeized gravitons  
using effective field theory methods similarly to the calculations carried out for the gluon's trajectory performed in \cite{Our,Our01}. Implying various issues investigated in
high energy gravity theories, see \cite{Veneziano,Veneziano1,Veneziano2,Veneziano3,Veneziano4,Veneziano5,Veneziano6} and references therein for the examples, we hope that the introduced calculation framework will allow to explore interesting problems in high energy gravity and beyond.

 The main result of the paper it is a recalculation of the reggeized graviton trajectory function and determination of the corresponding propagator of the reggeized gravitons, 
see \eq{Sca20001} and \eq{Sca21} expressions. The corresponding scattering amplitude calculated, see \eq{Ampl9} or \eq{Ampl15}, is different from the amplitude at a low-energy limit, i.e. from the ordinary static gravity results, see \cite{Donoghue,Donoghue1,Donoghue2,Donoghue3}. The main difference, if do not account the difference in the kinematics of high and low energy processes, is that in the high energy limit the amplitude is reggeized, see \eq{Sca21}. Nevertheless, it is hard to say if the reggeization of the graviton can be observed in any experiment of now days. Namely, talking about 
an inclusive quantities, total cross section for example, we now that the leading contribution to the inclusive amplitude at high energy is by eikonal amplitudes, 
see for the gravity sector calculation of \cite{GravExpon4,GravExpon5,GravExpon6,Veneziano,Veneziano1,Veneziano2,Veneziano3,Veneziano4,Veneziano5,Veneziano6}. In these quantities, the slope of the trajectory which is proportional to $q^{2}$, perhaps can not be of any experimental importance.  Still, the situation 
in QCD and phenomenological description of hadrons scattering is different, see for example \cite{MyPomeron} and references therein. It is well known,
the non-zero slope of any reggeized t-channel amplitude is especially important when we talk about less inclusive processes. All differential cross sections, production or difractive amplitudes depend on the
single t-channel one particle amplitude and there the reggeization with corresponding slope can be important. 
Of course, the situation in the experimental gravity very far from we have in QCD scattering experiments, but still there are a plenty theoretical calculations where the proper amplitude's trajectory must be properly accounted.

 Another interesting issue concerns the eikonal amplitudes, is a matter of one particle t-channel amplitude arises in the eikonal scattering. As we know, 
see for example \cite{MyPomeron,HighE}, the eikonal amplitude consists two main ingredients, amplitude and impact factors. 
The first one it is a t-channel one particle interacting amplitude which appears in the exponential.
The construction of this amplitude it is a complicated task, we must consider an interaction of scattering particles by one t-channel amplitude plus interaction by two t-channel amplitudes and so on, at the end usually an exponentiation of this one-particle amplitude introduced initially appears as LO answer.  
These calculations, see it for gravity sector in \cite{Veneziano3,Veneziano4,Veneziano5,Veneziano6} for example, 
are based on the unitarity requests and some perturbative schemes. Of course, as the initial t-channel amplitude we can use the reggeized graviton as well,
namely we can eikonalize the reggeized graviton reproducing known LO results plus some corrections. The corrections 
are definitely small because depend on the $q^{2}$ value but this is a normal situation for the eikonal scattering. It is well known that the t-channel amplitude's trajectory is still necessary 
and very important if we try to describe an experimental data, see \cite{MyPomeron,MyPomeron1} for example. So, even in the eikonalized amplitude we have to know the particle's trajectory if we wish to describe scattering data which we yet have no enough in the gravity sector unfortunately.
In any case,  in this context it is important what kind of amplitude appears in the exponential.
In high energy hadron scattering, for example, the t-channel amplitude 
is a phenomenological Pomeron, see \cite{MyPomeron} and not a Reggeon. At the same way, if we will talk about similar construction in QCD it must be a BFKL Pomeron in one or another form, see for example \cite{MyPomeron1} and references therein, and whole construction must be based on the interacting BFKL Pomeron scheme. The reason for the Pomeron and not Reggeon appearance is simple. A Pomeron is a bound state of two Reggeons and account LO s-channel inelastic production processes which are important
at high energy scattering, i.e. this construction is very important from the point of view of unitarity even when we talk about the inclusive scattering. Turning back to gravity, 
we realize that here we have a problem, we do not have a data and we do not know really what a gravitational Pomeron defined as a bound state of two reggeized gravitons is really. Therefore, the introduction of the reggeized graviton is a clarification of the possible unitarity corrections construction. It does not provides much if we talk about the eikonal even if we eikonalize the
Reggeized graviton, but it definitely important if we discuss an issue of the construction of gravity Pomeron and all related unitarity corrections.

 An additional issue concerns the approach is that we consider an effective field theory for the gravitational interactions valid at high energy limit
and based on BFKL physics ideas. In this extent, we construct a framework which allows to 
investigate different complex questions of high energy quantum gravity, see \cite{Veneziano,Veneziano1,Veneziano2,Veneziano3,Veneziano4,Veneziano5,Veneziano6}, by the calculations established in the 
effective quantum field theory. Namely, the principal structure of the approach is similar to the one of high energy QCD, see \cite{LipatovDL,Venug,Venug1,Venug2} for example, and we can 
to think about an application of the high energy QCD ideas in effective quantum gravity processes. 
For example, in QCD counterpart of Lipatov's effective action, we have a 
very fruitful BK (Balitsky-Kovchegov) equation approach, see \cite{Balitsky,Balitsky1,Kovchegov,OurBK}. It is interesting to explore, what can be analog of that type of scattering on gravity interactions side. 

 An another potentially interesting subject is about a clarification of the BFKL type of dynamics in an physical processes. In QCD, see for examples 
\cite{ExpQCD,ExpQCD1,ExpQCD2,ExpQCD3,ExpQCD4,ExpQCD5,ExpQCD6}
and references therein, a large amount of work has been invented to elucidate a BFKL dynamics in different types of scattering and particles production
processes. The reason for the 
complexity of this research is that the BFKL effect is a perturbative one and we have no many pure perturbative probes in high energy QCD. For example, having in hands a QCD BFKL Pomeron, see 
\cite{BFKL,BFKL1,BFKL2}, in the phenomenological description of  inclusive  cross sections we observe a phenomenological Pomeron, see \cite{Gribov, MyPomeron} with intercept drastically different from the BFKL one. The smallness of the intercept of the phenomenological Pomeron in comparison to the perturbative is caused, of course, by the plenty of unitary QCD corrections, which partially are non-perturbative ones. In gravity, in turn, the situation could be different. First of all, in orthodox gravity we have no a running coupling constant, it fixed, that makes the probes much clearer in principal. Secondly, the gravity coupling constant is small itself, so we can hope that the unitary corrections will not be such important as they do in QCD. Also, the gravity is one loop renormalizable, so we can hope also that
the answer will be finite. In this case, of course, the main matter is to find an observable in macroscopic classical gravity which can be sensitive to the energy regime of the microscopic process. It is also interesting to note, that in principle the behavior of the Pomeron amplitude is sensitive to the additional QFT contributions to the intercept arise from the supersymmetric theories, see \cite{LipatovGrav,LipatovGrav01,LipatovGrav02,LipatovGrav03,LipatovGrav04,LipatovGrav05,LipatovGrav06}. Therefore, if the microscopic quantum dynamics of the amplitude is affected by the additional degrees of freedom it can be interesting to analyze if we can observe this effect in macroscopic gravitational events.
Concluding we hope, that the use of the Lipatov's effective action approach and ideas of BFKL physics in high energy gravity can lead to the new interesting developments in the quantum theory
of gravity.
	
	The author is indebted to A. Sabio Vera for very helpful comments concern the subject of the paper.

\appendix
\newpage
\section{ Classical shock wave solution}\label{AppA}
\renewcommand{\theequation}{A.\arabic{equation}}
\setcounter{equation}{0}

 Let's consider the \eq{CSW18} form of the solution of  \eq{CSW5}. Firstly, we interesting to define a form of $\Phi$ function. 
Assuming that our field is symmetrical with respect to the polar angle, we write:
\beq\label{V1}
\Phi(x_{\bot})\,=\,f(r) \phi(\theta)\,.
\eeq
Correspondingly equations we solve are
\beqar\label{V2}
&\,&\frac{d^2 \phi}{d \theta^2}\,+\,n^2 \phi\,=\,0\,,\,\,\,\,n\,=\,1,\,2\,\cdots
\\
&\,&
r^2\,\frac{d^2 f}{dr^2}\,+\,r\,\frac{d f}{d r}\,-\,\Le m^2 r^2\,+\,n^2\Ra f\,=\,0\,.
\label{V3}
\eeqar
Here we excluded $n\,=\,0$ solution which corresponds to growing with $r$ function.
The solution of the first equation is trivial:
\beq\label{V4}
\phi(\theta)\,=\,\sum_{n\,=\,1}\,A_{n}\cos(n\,\theta\,+\,\alpha_{n})\,.
\eeq
The second equation is a Bessel equation for the complex argument, as it's solutions we choose 
a modified Hankel function or first kind (Hankel function of imaginary argument):
\beq\label{V5}
f(r)\,=\,H_{n}(\imath\, r m)\,,
\eeq
these functions at arbitrary $n$ are limited at the $r\,\rightarrow\,\infty$ limit, whereas  the second independent solution
(modified Bessel function of first kind) of the equation does not. Now, requiring that the solution will be real, instead the $H_{n}$ function we can define a real McDonald function as our solution:
\beq\label{V6}
f(r)\,=\,H_{n}(\imath\, r m)\,=\,\frac{2}{\pi}\,e^{-\imath\pi \Le n+1   \Ra/2}\,K_{n}(rm)\,,\,\,\,\,n\,=\,2k\,-\,1\,,\,\,\,k\,=\,1,2,\cdots\,.
\eeq
So, the solution for  our $\mA$ field we obtain is:
\beq\label{V7}
\mA(x^{-},x_{\bot})\,=\,\frac{2}{\pi}\,Z(x^{-})\,\sum_{n\,=\,1}\,A_{n} K_{n}(rm)\cos(n\,\theta\,+\,\alpha_{n})\,
\eeq
or in more general form
\beq\label{V701}
\mA(x^{-},x_{\bot})\,=\,\frac{2}{\pi}\,\sum_{n\,=\,1}\,A_{n}(x^{-}) K_{n}(rm)\cos(n\,\theta\,+\,\alpha_{n})\,
\eeq
Next step we have is to apply the initial conditions in our problem set-up. For simplicity we assume that the initial for of the wave is given by the 
first harmonic ($n\,=\,1$) only with zero initial phase, we write correspondingly:
\beq\label{V8}
\mA(x^{-},x_{\bot})\,=\,\frac{2}{\pi}\,Z(x^{-})\,A_{1}\, K_{1}(rm)\cos(\theta)\,.
\eeq
The $Z(x^{-})$ function can be found from the initial density of the field in turn:
\beq\label{V9}
\,T^{++}\,\propto\,\Le \D_{-} Z(x^{-})\Ra^{2}\,.
\eeq
We do no consider a precise solution of this problem here, the precise solutions requires an precise definition of the initial condition for the
$\mA$ field but it lies above the scope of the paper. Also we note that for our aims the \eq{V8} expression is enough.
Additional remark is about the $r$ dependence of the field. As usual, the solution does not work at the $r\,\rightarrow\,0$ limit, as any classical solution.
Moreover this limit is beyond the weak field approximation we use of course. Therefore, roughly speaking, we can apply the solution only at $r\,\geq\,1/m$ limit
where the $K_{1}$ function can be expanded in an asymptotic series.

\newpage
\section{ Next order corrections to the impact factor}\label{AppB}
\renewcommand{\theequation}{B.\arabic{equation}}
\setcounter{equation}{0}

 In this Appendix we present a calculation of the impact factor by accounting the gravitational field fluctuation as weel, we continue here the calculation of the \ref{ImpFac} Section.
So, our next step, is an accounting of the terms linear with respect to $\epsilon_{\mu \nu}\,$ in the \eq{effa17} effective action, they  are: 
\beqar
&\,&S_{eff\,1}(y\,=\,0)=\int d^{4} x\Le\,-\,
\,\int\,d^{4}y\,V_{++\,1}\,\Le \D_{-}^{2}\phi_{f}\,+\,\D_{-}^{2}\mA\Ra\,G(x,y,\epsilon=0)\,V_{++\,0}\,\Le \D_{-}^{2}\phi_{f}\,+\,\D_{-}^{2}\mA\Ra\,
\,-\,
\right.\nonumber \\
&-&\,\left.
\,\int\,d^{4}y\,\Le V_{++\,0}\,\Le \D_{-}^{2}\phi_{f}\,+\,\D_{-}^{2}\mA\Ra\Ra_{x}\,G(x,y,\epsilon=0)\,\Le \epsilon_{--}\,\D_{+}^{2}\phi_{f}\Ra_{y}\,-\,
\frac{1}{2}\,\epsilon_{--}\,\D_{+}\phi_{f} \,\D_{+}\phi_{f}\,-\,
\right.\nonumber \\
&-&\,\left.
\frac{1}{2}\,\int\,d^{4}y\,\Le V_{++\,0}\,\Le \D_{-}^{2}\phi_{f}\,+\,\D_{-}^{2}\mA\Ra\Ra_{x}\,\,G_{1}(x,y)\,
\Le V_{++\,0}\,\Le \D_{-}^{2}\phi_{f}\,+\,\D_{-}^{2}\mA\Ra\Ra_{y}\,-\,
\right.\nonumber \\
&-&\,\left.
\frac{1}{2}\,V_{++\,1}\Le 2\D_{-}\phi_{f}\,\D_{-}\mA+\D_{-}\phi_{f}\,\D_{-}\phi_{f}\Ra\Ra\,+\,
\frac{\imath}{2}\, Sp\,\Le\ln\Le 1\,+\,G_{0}\Le V_{++}\,\D_{-}^{2}\,+\,\epsilon_{--}\,\D_{+}^{2} \Ra  \Ra\Ra_{1}\,\label{effa1703}.
\eeqar
where $G_{1}(x,y)$ denotes the full propagator of scalar field linear with respect to the gravitational field fluctuations
see \eq{effa16} expression. The same denotes the 
subscript on the vacuum loop expression, i.e. the subscript of the \eq{effa1703} last term which means that in the perturbative expansion we have to account the 
only linear with respect to $\epsilon_{\mu \nu}$ terms. Correspondingly
\beq\label{effa1704}
V_{++\,1}\,=\,\epsilon_{++}+\frac{1}{2}\Le \D_{i}\D_{+}^{-1}\epsilon_{++} - 2\epsilon_{+i}\Ra\,\Le \D_{i}\D_{+}^{-1}\Le \mB_{++}+h_{f\,++}\Ra \Ra\,.
\eeq
is a linear interaction vertex.

  Next we consider the terms which
quadratic with respect to the fluctuations in the way determined by the \eq{effa12} Lagrangian,
i.e. terms which determine the propagator of the gravitational field and additional loop corrections to the impact factor.  From the functional determinant we preserve only the two first terms of it's perturbative expansion:
\beqar\label{effa18}
&\,&\,\frac{\imath}{2}\,Sp\,\ln\Le 1\,+\,G_{0}\Le V_{++}\,\D_{-}^{2}\,+\,\epsilon_{--}\,\D_{+}^{2} \Ra  \Ra\,\rightarrow\,
\\
&\rightarrow&\,
\,\frac{\imath}{2}\,Sp\Le G_{0}\Le V_{++}\,\D_{-}^{2}\,+\,\epsilon_{--}\,\D_{+}^{2} \Ra\Ra\,-\,
\frac{\imath}{4}\,Sp\Le G_{0}\Le V_{++}\,\D_{-}^{2}\,+\,\epsilon_{--}\,\D_{+}^{2}\Ra\,G_{0}\,\Le V_{++}\,\D_{-}^{2}\,+\,\epsilon_{--}\,\D_{+}^{2} \Ra\Ra\,.
\nonumber
\eeqar
Correspondingly, the Lagrangian's terms which determine the graviton propagator are:
\beqar\label{effa19}
&\,& L_{eff\,pr}(y\,=\,0)\,=\,
\frac{1}{2\kappa}\,\epsilon_{--}\,\Box\,\epsilon_{++}\,-\,
\frac{1}{\kappa}\,\epsilon_{+i}\,\Box\,\epsilon_{-i}\,-\,
\,\int\,d^{4}y\,\Le\epsilon_{++}\,\Le \D_{-}^{2}\phi_{f}\,+\,\D_{-}^{2}\mA\Ra\Ra_{x}\,G_{0}(x,y)\,\Le \epsilon_{--}\,\D_{+}^{2}\phi_{f}\Ra_{y}\,-\,
\nonumber \\
&-&\,
\int\,d^{4}y\,\Le\Le\D_{i}\D_{+}^{-1}\epsilon_{++}\Ra \Le \epsilon_{+i}\,+\,\frac{1}{2}\D_{i}\D_{+}^{-1}\Le \mB_{++}+h_{f\,++} \Ra\Ra\,
\Le \D_{-}^{2}\phi_{f}\,+\,\D_{-}^{2}\mA\Ra\Ra_{x}\,G_{0}(x,y)\,\Le \epsilon_{--}\,\D_{+}^{2}\phi_{f}\Ra_{y}\,
\,-\, 
\nonumber \\
&-&
\frac{\imath}{2}\,\int\,d^{4} y\,
G_{0}(x,y)\Le \epsilon_{++}\,\D_{-}^{2}\Ra_{y}\,G_{0}(y,x)\,\Le \epsilon_{--}\,\D_{+}^{2} \Ra_{x}\,-\,
\nonumber\\
&-&
\frac{\imath}{2}\,\int\,d^{4} y\,
G_{0}(x,y)
\Le \Le\D_{i}\D_{+}^{-1}\epsilon_{++}\Ra \Le \epsilon_{+i}\,+\,\frac{1}{2}\D_{i}\D_{+}^{-1}\Le \mB_{++}+h_{f\,++} \Ra\Ra\,\D_{-}^{2}\Ra_{y}\,
G_{0}(y,x)\,\Le \epsilon_{--}\,\D_{+}^{2} \Ra_{x}\,.
\eeqar
There are the terms which are quadratic with respect to fluctuations but provide the sub-leading contributions (non-diagonal ) to the propagators. These terms are:
\beqar
&\,& S_{eff\,2}(y\,=\,0)\,=\,\int d^{4} x\Le\,
\frac{1}{2\kappa}\,\Le  -\frac{1}{2}\D_{i}\D_{-}^{-1}\epsilon_{--}\,+\,\epsilon_{-i}\Ra^2\,\D_{i}^{2} \mB_{++}
\,-\,
\right.\nonumber \\
&-&\,\left.
\frac{1}{2}\,
\,\int\,d^{4}y\,V_{++\,1}\,\Le \D_{-}^{2}\phi_{f}\,+\,\D_{-}^{2}\mA\Ra\,G(x,y,\epsilon=0)\,V_{++\,1}\,\Le \D_{-}^{2}\phi_{f}\,+\,\D_{-}^{2}\mA\Ra\,
\,-\,
\right.\nonumber \\
&-&\,\left.
\,\int\,d^{4}y\,V_{++\,2}\,\Le \D_{-}^{2}\phi_{f}\,+\,\D_{-}^{2}\mA\Ra\,G(x,y,\epsilon=0)\,V_{++\,0}\,\Le \D_{-}^{2}\phi_{f}\,+\,\D_{-}^{2}\mA\Ra\,-\,
\right.\nonumber \\
&-&\,\left.
\,\int\,d^{4}y\,V_{++\,1}\,\Le \D_{-}^{2}\phi_{f}\,+\,\D_{-}^{2}\mA\Ra\,G_{1}(x,y)\,V_{++\,0}\,\Le \D_{-}^{2}\phi_{f}\,+\,\D_{-}^{2}\mA\Ra\,-\,
\right.\nonumber \\
&-&\,\left.
\,\frac{1}{2}\,\int\,d^{4}y\,V_{++\,0}\,\Le \D_{-}^{2}\phi_{f}\,+\,\D_{-}^{2}\mA\Ra\,G_{2}(x,y)\,V_{++\,0}\,\Le \D_{-}^{2}\phi_{f}\,+\,\D_{-}^{2}\mA\Ra\,-\,
\right.\nonumber \\
&-&\,\left.
\frac{1}{2}\,\,\int\,d^{4}y\,\epsilon_{--}\,\D_{+}^{2}\phi_{f}\,G(x,y,\epsilon=0)\,\epsilon_{--}\,\D_{+}^{2}\phi_{f}\,-\,
\right.\nonumber \\
&-&\,\left.
\int\,d^{4}y\, V_{++\,0}\,\Le \D_{-}^{2}\phi_{f}\,+\,\D_{-}^{2}\mA\Ra\,G_{1}(x,y,\epsilon_{++}=0)\,\epsilon_{--}\,\D_{+}^{2}\phi_{f}\,-\,
\right.\nonumber \\
&-&\,\left.
\int\,d^{4}y\, V_{++\,1}(\epsilon_{++}=0)\,\Le \D_{-}^{2}\phi_{f}\,+\,\D_{-}^{2}\mA\Ra\,G(x,y,\epsilon=0)\,\epsilon_{--}\,\D_{+}^{2}\phi_{f}\,-\,
\right.\nonumber \\
&-&\,\left.
\frac{1}{2}\,V_{++\,2}\Le 2\D_{-}\phi_{f}\,\D_{-}\mA+\D_{-}\phi_{f}\,\D_{-}\phi_{f}\Ra\Ra\,+\,
\frac{\imath}{2}\, Sp\,\Le\ln\Le 1\,+\,G_{0}\Le V_{++}\,\D_{-}^{2}\,+\,\epsilon_{--}\,\D_{+}^{2} \Ra  \Ra\Ra_{2}\,,\label{effa20}
\eeqar
the terms accounted already in \eq{effa19} should not be taken into account in this expression of course.
Here
\beq\label{effa21}
V_{++\,2}\,=\,\frac{1}{4}\,\Le \D_{i}\D_{+}^{-1}\epsilon_{++}\Ra^2\,-\,\frac{1}{2}\,\epsilon_{+i}\,\Le \D_{i}\D_{+}^{-1}\epsilon_{++}\Ra\,+\,\epsilon_{+i}^{2}\,,
\eeq
see again \eq{effa14}-\eq{effa1401} expression.
This part of the action provides some corrections to the diagonal propagators of the problem, we do not discuss these terms further.

 Now we return to the integration over the gravitational fluctuations. The definition of propagators we use is a standard one:
\beq\label{effa22}
\frac{\delta^{2} S_{eff\,pr}}{\delta\epsilon_{\mu \nu}\delta\epsilon_{\rho\sigma}}\,G_{\rho \sigma \alpha \beta}\,=\,
S^{\mu \nu \rho \sigma}\,G_{\rho \sigma \alpha \beta}\,=\,
\frac{1}{2}\,\Le \delta^{\mu}_{\alpha}\delta^{\nu}_{\beta}\,+\,\delta^{\mu}_{\beta}\delta^{\nu}_{\alpha}\Ra\,.
\eeq
In the case of bare propagators we need, we have correspondingly:
\beq\label{effa22002}
S^{--++}\,G_{0\,++--}\,=\,\frac{1}{2\kappa}\,\Box\,G_{0\,++--}\,=\,1\,,\,\,\,\,G_{0\,++--}\,=\,-\,2\kappa\,\int\,\frac{d^{4}p}{(2\pi)^{4}}\,\frac{e^{-\imath\,p\,(x-y)}}{p^{2}\,+\,\imath\varepsilon}\,;
\eeq
and
\beq\label{effa23}
S^{+i-k}\,G_{0\,-k+j}\,=\,\frac{1}{\kappa}\,\eta^{ik}\Box\,G_{-k+j}\,=\,\frac{1}{2}\,\delta^{i}_{j}\,,\,\,\,\,
G_{0\,-i+j}\,=\,-\,\frac{\kappa}{2}\,\eta_{ij}\,\int\,\frac{d^{4}p}{(2\pi)^{4}}\,\frac{e^{-\imath\,p\,(x-y)}}{p^{2}\,+\,\imath\varepsilon}\,.
\eeq
Whereas the $G_{0\,-k+j}$ propagator has no corrections appear in \eq{effa19}, the full  $G_{++--}$ propagator is determined by the following expression\footnote{We do not include here the corrections
arise from the \eq{effa20} part of th eeffective action.} 
\beq\label{effa24}
\Le \frac{1}{2\kappa}\,\Box\,+\,M\,\Ra\,G_{++--}\,=\,\,1\,
\eeq
with corresponding solution
\beq\label{effa25}
G_{++--}(x,y)\,=\,G_{0\,x,y;++--}(x,y)\,-\,\int\,d^4 z\,d^4 z_1\,G_{0\,++--}(x,z)\,M(z,z_1)\,G_{++--}(z_1,y)\,
\eeq
where
\beqar\label{effa26}
M(z,z_1)\,&=&\,-\,\Le \D_{-}^{2}\phi_{f}\,+\,\D_{-}^{2}\mA\Ra_{z}\,G_{0}(z,z_1)\,\Le \D_{+}^{2}\phi_{f}\Ra_{z_1}\,-\,
\frac{\imath}{2}\,\D_{+z}^{2}\D_{-z_1}^{2}\left[G_{0}(z,z_1)\,G_{0}(z_1,z)\,\right]\,-\,
\\
&-&\,
\D_{iz}\D_{+z}^{-1}\,\left[\Le\Le \epsilon_{+i}\,+\,\frac{1}{2}\D_{i}\D_{+}^{-1}\Le \mB_{++}+h_{f\,++} \Ra\Ra\,
\Le \D_{-}^{2}\phi_{f}\,+\,\D_{-}^{2}\mA\Ra\Ra_{z}\,G_{0}(z,z_1)\,\Le \D_{+}^{2}\phi_{f}\Ra_{z_1}\right]\,-\,
\nonumber \\
&-&\,
\frac{\imath}{2}\,\D_{iz_1}\D_{+z_1}^{-1}\,\left[
\Le \epsilon_{+i}\,+\,\frac{1}{2}\D_{i}\D_{+}^{-1}\Le \mB_{++}+h_{f\,++} \Ra\Ra_{z_1}\,
\D_{-z_1}^{2}\,\D_{+z}^{2}\,\Le G_{0}(z_1,z)\,G_{0}(z,z_1)\Ra\right]\,
\nonumber
\eeqar
is a non-local vertex calculated till one scalar loop precision, i.e. a self-energy correction to the graviton's propagator. 

 Therefore, the only bare (no-loop) terms of the impact factor answer we have are:
\beq\label{effa27}
S_{eff}(y\,=\,0)\,=\,S_{eff\,0}(y\,=\,0)\,+\,S_{eff\,1}(y\,=\,0)
\eeq
with
\beqar
&\,&S_{eff\,0}(y\,=\,0)=\int d^{4} x\Le\,-\,
\frac{1}{2}\,\int\,d^{4}y\,V_{++\,0}\,\Le \D_{-}^{2}\phi_{f}\,+\,\D_{-}^{2}\mA\Ra\,G(x,y,\epsilon=0)\,
V_{++\,0}\,\Le \D_{-}^{2}\phi_{f}\,+\,\D_{-}^{2}\mA\Ra\,
\,-\,
\right.\nonumber \\
&-&\,\left.
\frac{1}{2}\,V_{++\,0}\Le 2\D_{-}\phi_{f}\,\D_{-}\mA+\D_{-}\phi_{f}\,\D_{-}\phi_{f}\Ra\Ra\,
\eeqar
and
\beq\label{effa29}
S_{eff\,1}(y\,=\,0)\,=\,-\,\kappa\,\int d^{4}x\,d^{4}y\,J_{++}(x)\,G_{--++}(x,y)\,J_{--}(y)\,+\,
\frac{\kappa}{2}\,\int d^{4}x\,d^{4}y\,J_{+i}(x)\,G_{-i+j}(x,y)\,J_{-j}(y)\,
\eeq
where
\beqar\label{effa30}
J_{--}(x)\,&=&\,
-\,\int\,d^{4}y\,\Le \D_{-}^{2}\phi_{f}\,+\,\D_{-}^{2}\mA\Ra_{x}\,G_{0}(x,y)\,\Le V_{++\,0}\,\Le \D_{-}^{2}\phi_{f}\,+\,\D_{-}^{2}\mA\Ra\Ra_{y}\,-\,
\nonumber \\
&-&\,
\frac{1}{2}\,\D_{ix}\D_{+x}^{-1}
\int\,d^{4}y\,\Le\Le \D_{i}\D_{+}^{-1}\Le \mB_{++}+h_{f\,++}\Ra\Ra_{x}\, 
\Le \D_{-}^{2}\phi_{f}\,+\,\D_{-}^{2}\mA\Ra_{x}\,G_{0}(x,y)\Ra\,\Le V_{++\,0}\,\Le \D_{-}^{2}\phi_{f}\,+\,\D_{-}^{2}\mA\Ra\Ra_{y}\,+\,
\nonumber \\
&+&\,
\frac{1}{2}\,\int d^4 y\,\int d^4 z\,
\Le V_{++\,0}\,\Le \D_{-}^{2}\phi_{f}\,+\,\D_{-}^{2}\mA\Ra\Ra_{z}\,
G_{0}(z,x)\,\D_{-x}^{2} G_{0}(x,y)\,\Le V_{++\,0}\,\Le \D_{-}^{2}\phi_{f}\,+\,\D_{-}^{2}\mA\Ra\Ra_{y}\,+\,
\nonumber \\
&+&\,
\frac{1}{4}\,\int d^4 y\,\int d^4 z\,
\Le V_{++\,0}\,\Le \D_{-}^{2}\phi_{f}\,+\,\D_{-}^{2}\mA\Ra\Ra_{z}\,
\D_{ix}\D_{+x}^{-1}\Le G_{0}(z,x)\,\Le\D_{i}\D_{+}^{-1}\Le \mB_{++}+h_{f\,++}\Ra\Ra_{x}\,\D_{-x}^{2}G_{0}(x,y)\Ra\,
\nonumber \\
&\,&
\Le V_{++\,0}\,\Le \D_{-}^{2}\phi_{f}\,+\,\D_{-}^{2}\mA\Ra\Ra_{y}\,-\,
\frac{1}{2}\,\Le 2\D_{-}\phi_{f}\,\D_{-}\mA+\D_{-}\phi_{f}\,\D_{-}\phi_{f}\Ra\,
\Le 1\,+\,\frac{1}{2} \D_{i}\D_{+}^{-1}\Le \mB_{++}+h_{f\,++}\Ra\Ra\,\nonumber
\eeqar
and correspondingly
\beqar\label{effa31}
J_{++}(x)\,&=&\,
\,\int\,d^{4}y\,\Le V_{++\,0}\,\Le \D_{-}^{2}\phi_{f}\,+\,\D_{-}^{2}\mA\Ra\Ra_{y}\,G(y,x,\epsilon=0)\,\D_{+x}^{2}\phi_{fx}\,+\,
\frac{1}{2}\,\D_{+}\phi_{f} \,\D_{+}\phi_{f}\,-\,
\nonumber \\
&+&\,
\frac{1}{2}\,\int d^4 y\,\int d^4 z\,
\Le V_{++\,0}\,\Le \D_{-}^{2}\phi_{f}\,+\,\D_{-}^{2}\mA\Ra\Ra_{z}\,
G_{0}(z,x)\,\D_{+x}^{2} G_{0}(x,y)\,\Le V_{++\,0}\,\Le \D_{-}^{2}\phi_{f}\,+\,\D_{-}^{2}\mA\Ra\Ra_{y}\,+\,
\nonumber \\
&+&\,
\frac{1}{2\kappa}\,\D_{i}\D_{-}^{-1}\Le \epsilon_{-i} \,\D_{i}^{2} \mB_{++}\Ra.
\eeqar
We have also
\beqar\label{effa33}
J_{+i}(x)\,&=&\,
\int\,d^{4}y\,
\Le \D_{i}\D_{+}^{-1}\Le \mB_{++}+h_{f\,++}\Ra\,\Le \D_{-}^{2}\phi_{f}\,+\,\D_{-}^{2}\mA\Ra\Ra_{x}\,G_{0}(x,y)\,\Le V_{++\,0}\,\Le \D_{-}^{2}\phi_{f}\,+\,\D_{-}^{2}\mA\Ra\Ra_{y}\,+\,
\nonumber \\
&+&\,\frac{1}{2}\,
\D_{i}\D_{+}^{-1}\Le \mB_{++}+h_{f\,++}\Ra\,\Le 2\D_{-}\phi_{f}\,\D_{-}\mA+\D_{-}\phi_{f}\,\D_{-}\phi_{f}\Ra\,
\eeqar
with $J_{-j}$ arises from the first term of \eq{effa29} with the last term of \eq{effa31} included. The obtained impact factor is given by \eq{effa27}-\eq{effa29} expressions, we also note that the second term of \eq{effa29}  is sub-leading and can be neglected if we consider the only leading contributions to the scattering amplitude.

\newpage
\section{ One loop corrections to the $G_{++--}$ propagator}\label{AppC}
\renewcommand{\theequation}{C.\arabic{equation}}
\setcounter{equation}{0}

 The \eq{Sca3} effective action structure provides  $S^{--}\,\propto\,\mB_{++}$, $S^{++}\,\propto\,\mB_{--}$, $S^{+-}_{1}\,\propto\,\mB$ and $S^{+-}_{2}\,\propto\,\mB_{--}\,\mB_{++}$.
For our purposes we need the only part of the full Green's function which remains after variations and consequent $\mB\rightarrow 0$ limit taken.
So, taking \eq{Sca8} full expression, the part of the propagator we need is:
\beqar\label{C1}
G_{+-}(x,y)\,&=&\,G_{0\,+-}(x,y)\,-\,G_{0\,+-}(x,z)\,S_{1}^{-+}(z,z_{1})\,G_{0\,+-}(z_{1},y)\,-\,G_{0\,+-}(x,z)\,S_{2}^{-+}(z,z_{1})\,G_{0\,+-}(z_{1},y)\,+\,
\\
&+&
G_{0\,+-}(x,z)\,S_{1}^{-+}(z,z_{1})\,G_{0\,+-}(z_{1},z_{2})\,S_{1}^{-+}(z_{2},z_{3})\,G_{0\,+-}(z_{3},y)\,+\,
\nonumber \\
&+&\,
G_{0\,+-}(x,z)\,S^{--}(z,z_1)\,G_{0\,-+}(z_1,z_2)\,S^{++}(z_2,z_3)\,G_{0\,+-}(z_3,y)\,\nonumber.
\eeqar
Next object we need are the vertices of the effective interactions. The first two vertices we have ( the second is obtained from the first one by the index sign change),  
we obtain by the functional derivative of the following type:
\beqar
S^{--}(z,w)\,& = &\,\frac{\delta^{2} S_{eff}}{\delta \epsilon_{-}(z)\,\delta \epsilon_{-}(w)}\,=\,\frac{1}{8\kappa}\,\frac{\delta^{2} }{\delta \epsilon_{-}(z)\,\delta \epsilon_{-}(w)}\,
\int\, d^4 x\,\D_{i}^{2} \mB_{+}\,\Le \D_{i}\D_{-}^{-1}\epsilon_{-}\Ra^{2}\,=\,
\nonumber \\
&=&\,
\frac{1}{8\kappa}\,\frac{\delta^{2} }{\delta \epsilon_{-}(z)\,\delta \epsilon_{-}(w)}\,
\int\, d^4 x\,\D_{i}^{2} \mB_{+}\,\Le \D_{i x}\int dx_{1}^{-}\,f(x^{-},x_{1}^{-})\epsilon_{-}(x_{1}^{-},x^{+},x_{\bot})\Ra^{2}\,\label{C2}.
\eeqar
Here, the $f(x^{-},x_{1}^{-})$ function is defined through a theta function, there are few possibilities for it's definition, we have for example: 
\beqar\label{C3}
f(x^{-},x_{1}^{-})\,& = &\,\theta(x^{-}\,-\,x_{1}^{-});
\\
f(x^{-},x_{1}^{-})\,& = &\,-\,\theta(x^{-}_{1}\,-\,x^{-});
\label{C31}\\
f(x^{-},x_{1}^{-})\,& = &\,\frac{1}{2}\,\Le\theta(x^{-}\,-\,x_{1}^{-})\,-\,\theta(x^{-}_{1}\,-\,x^{-})\Ra\,\label{C32}.
\eeqar
Now we have to clarify how to understand the square of the $\D_{\pm}^{-1}$ operator in all expressions. The clue here comes from the fact, that in QCD the effective currents of the effective action
represent ordered in rapidity space fields in the form of ordered exponentials, see \cite{Our,Our01}. This ordering of interacting fields is an important part of the BFKL kernel construction in general
of course. 
So, our requirement is that the square of two fields, which in our case are 
$\epsilon_{\pm}$, must be not local but ordered in rapidity space. Formally it means that we have to define the square of the two fields in the effective current as :
\beq\label{C3201}
\Le \D_{-}^{-1}\epsilon_{-}\Ra^{2}\,=\,-\,\int^{\infty}_{y^{-}} dy_{1}^{-}\,\int_{-\infty}^{y^{-}} dy_{2}^{-}\,\epsilon_{-}(y_{1}^{-})\,\epsilon_{-}(y_{2}^{-})\,
\eeq
in the simplest case of $f$ functions. The uncertainty, of course, is a result of the perturbative construction of the effective current. Whereas in QCD we know the form of the currents as whole, here we calculate it order by order loosing part of the important information in the calculations. The check of the correctness of the expression in fact is simple. First of all, we can calculate the reggeized propagator and notice that only this
form of the squared operator provides an answer which is not singular and independent on the way of the contour integration, see further. An another check, it is a comparison of the final answer with the 
answer from  \cite{LipatovGrav,LipatovGrav01,LipatovGrav02,LipatovGrav03,LipatovGrav04,LipatovGrav05,LipatovGrav06} papers, where the position of the poles of the amplitude is present.
Taking derivatives we obtain finally:
\beq\label{C33}
S^{--}(z,w)\,= \,-\,\frac{1}{4\kappa}\,\D_{iz}^{2} \mB_{+}(z)\,\int\,dy^{-}\,f(w^{-},y^{-})\,f(y^{-},z^{-})\,
\delta(z^{+}-w^{+})\,\delta^{2}(z_{\bot}-w_{\bot})\,\D_{iw}^{2}\,.
\eeq
The particular ordering of the arguments in the expression is related to the ordering of "in" and "out" Fourier components in the corresponding Green's function
in rapidity space, i.e.
to the direction of the ordered exponential aka Wilson line. The second argument, which is $w$ in this case, is an argument of "in" plane wave whereas the first
argument,  $z$, is an argument of "out" wave that determines the form of the $f$ functions.
Correspondingly we have:
\beq
\frac{\delta S^{--}(z,w)}{\delta \mB_{+}(x)}\,=\,-\,
\frac{1}{4\kappa}\,
\int\,dy^{-}\,f(w^{-},y^{-})\,f(y^{-},z^{-})\,\,
\delta(z^{+}-w^{+})\,\delta^{2}(z_{\bot}-w_{\bot})\,\delta(z^{+}-x^{+})\,\delta^{2}(z_{\bot}-x_{\bot})\,\D_{jx}^{2}\,
\Le \D_{iz}\,\D_{iw}\Ra\,\label{C4}.
\eeq
In the expression we precisely wrote the $\D_{iz}\,\D_{iw}$ derivatives in the symmetrical form understanding the action of the operator as firstly taking these derivatives from the two different
Green's function in an expression for the kernel and only after that performing an integration over the transverse coordinates, see further.
The second type of the vertex we need is the  $S^{+-}$ which has linear and quadratic with respect to the reggeized graviton field terms. We have: 
\beq\label{C5}
S^{+-}(z,w)\,=\,\frac{\delta^{2} S_{eff}}{\delta \epsilon_{+}(z)\,\delta \epsilon_{-}(w)}\,,
\eeq
the shortened notatins for the fluctuation fields are used here again.
For the vertex's calculation we need
the following NLO parts of the full effective action:
\beqar\label{C6}
S_{eff}^{1}& = &
\frac{1}{4\kappa}\,\int\, d^4 x\,\Le \epsilon_{-}\,\Le
\D_{i}^{2}\,\mB_{+}\,\D_{j}^{2}\,\D_{+}^{-1}\,\D_{+}^{-1}\Ra\,\epsilon_{+}\,+\,
\epsilon_{+}\,\Le \D_{j}^{2}\,\D_{-}^{-1}\,\D_{-}^{-1}\,\D_{i}^{2}\,\mB_{-}\,
\Ra\,\epsilon_{-}\,\Ra\,-\,
 \\
&-&\,
\frac{1}{2\kappa^2}\,
\int\, d^4 x\,\int d^4 y\,\left[\epsilon_{-}\,\Le \D_{i}^{2} \mB_{+}\, \D_{j}\,\D_{-}^{-1}\Ra\right]_{x}\,
\left[ \epsilon_{+}\,\Le \D_{i}^{2} \mB_{-}\, \D_{k}\,\D_{+}^{-1}\Ra\right]_{y}\,G_{0\,-j+k}(x,y)\,
\nonumber
\eeqar
and for the linear vertex we obtain:
\beqar\label{C7}
S^{+-}_{1}(z,w)\,& = &\,\frac{1}{4\kappa}\,\D_{i w}^{2}\,\mB_{+}(w)\,\int\,dy^{+}\,f(w^{+},y^{+})\,f(y^{+},z^{+})\,\delta(z^{-}-w^{-})\,\delta^{2}(z_{\bot}-w_{\bot})\,\D_{iz}^{2}\,+\,
\\
&+&\,\frac{1}{4\kappa}\,
\D_{i z}^{2}\,\mB_{-}(z)\,\int\,dy^{-}\,f(z^{-},y^{-})\,f(y^{-},w^{-})\,\delta(z^{+}-w^{+})\,\delta^{2}(z_{\bot}-w_{\bot})\,\D_{iw}^{2}\,
\nonumber.
\eeqar
Correspondingly:
\beq\label{C8}
\frac{\delta S^{+-}_{1}(z,w)}{\delta \mB_{+}(x)}\,=\,\frac{1}{4\kappa}\,
\int\,dy^{+}\,f(w^{+},y^{+})\,f(y^{+},z^{+})\,\delta(z^{-}-w^{-})\,\delta(x^{+}-w^{+})\,\delta^{2}(z_{\bot}-w_{\bot})\,
\delta^{2}(x_{\bot}-w_{\bot})\,
\D_{ix}^{2}\,\D_{iz}^{2}\,
\eeq
and 
\beq\label{C9}
\frac{\delta S^{+-}_{1}(z,w)}{\delta \mB_{-}(x)}\,=\,\frac{1}{4\kappa}\,
\int\,dy^{-}\,f(z^{-},y^{-})\,f(y^{-},w^{-})\,\delta(z^{+}-w^{+})\,\delta(z^{-}-x^{-})\,\delta^{2}(z_{\bot}-w_{\bot})\,\delta^{2}(x_{\bot}-z_{\bot})\,
\D_{iy}^{2}\,\D_{iw}^{2}\,\,
\eeq
The vertex quadratic with respect to the reggeized gravitons is:
\beqar\label{C10}
S^{+-}_{2}(z,w)\, & = &\,-\,
\frac{1}{2\kappa^2}\,
\left[\Le \D_{i}^{2} \mB_{+}\, \D_{j}\,\D_{-}^{-1}\Ra\right]_{w}\,
\left[\Le \D_{i}^{2} \mB_{-}\, \D_{k}\,\D_{+}^{-1}\Ra\right]_{z}\,G_{0\,-j+k}(w,z)\,=\,
\\
&=&\,-\,
\frac{1}{2\kappa^2}\,
\left[\D_{i}^{2} \mB_{+}\, \D_{j} \right]_{w}\,\left[\D_{i}^{2} \mB_{-}\, \D_{k}\right]_{z}\,G_{0\,-j+k}(w,z)\,
\int\,ds^{-}\,f(s^{-},w^{-})\,\int\,dt^{+}\,f(t^{+},z^{+})\,
\nonumber
\eeqar
that gives after the variation with respect to both graviton's fields:
\beqar\label{C11}
\frac{\delta^{2} S^{+-}_{2}(z,w)}{\delta \mB_{+}(x)\delta \mB_{-}(y)}& = &-\frac{1}{2\kappa^2}
\left[\D_{i}^{2} \, \D_{j}\right]_{w}
\left[\D_{i}^{2} \, \D_{k}\right]_{z}
G_{0\,-j+k}(w,z)
\\
&\,&
\int ds^{-}\,f(s^{-},w^{-})\int dt^{+}\,f(t^{+},z^{+})\,
\delta(x^{+}-w^{+})\delta(y^{-}-z^{-})\delta^{2}(x_{\bot}-w_{\bot})\delta^{2}(y_{\bot}-z_{\bot})\,.
\eeqar
The vertices determine the structure of the one-loop correction to the propagator.
\,\newline

 The kernel of the correction is defined through the simple variations:
\beq\label{C12}
2\,\imath\, S_{xy\,1}=\Le\frac{\delta^2\,Sp\ln(G)}{\delta\mB_{+x}\,\delta \mB_{-y}}\Ra_{\mB_{\pm}=0}=\frac{\delta^2\,G_{+-}(z,z_1)}{\delta\mB_{+x}\,\delta \mB_{-y}}\, G_{+-}^{-1}(z_1,z)-
\frac{\delta\,G_{+-}(z,z_1)}{\delta \mB_{-y}}\, G_{+-}^{-1}(z_1,z_2)\,\frac{\delta\,G_{+-}(z_2,z_3)}{\delta \mB_{+x}}\, G_{+-}^{-1}(z_3,z)
\eeq
with Green's function given by \eq{C1} expression. We have then for the first term in the expression:
\beqar\label{C13}
2\,\imath\, S_{xy\,1\,1}\,& = &\,\frac{\delta^2\,G_{+-}(z,z_1)}{\delta\mB_{+x}\,\delta \mB_{-y}}\, G_{+-}^{-1}(z_1,z) \,=\, -\,G_{0\,+-}(z,w)\,\frac{\delta^{2} S^{-+}_{2}(w,z)}{\delta \mB_{+x}\delta \mB_{-y}}\,+\,
\nonumber \\
&+&
G_{0\,+-}(z,w)\,\frac{\delta S^{--}(w,w_1)}{\delta \mB_{+x}}\,G_{0\,-+}(w_1,w_2)\,\frac{\delta S^{++}(w_2,z)}{\delta \mB_{-y}}\,+\,
\nonumber \\
&+&
G_{0\,+-}(z,w)\,\frac{\delta S_{1}^{-+}(w,w_{1})}{\delta\mB_{+x}}\,G_{0\,+-}(w_{1},w_{2})\,\frac{\delta  S_{1}^{-+}(w_{2},z)}{\delta\mB_{-y}}\,+\,
\nonumber \\
&+&
G_{0\,+-}(z,w)\,\frac{\delta S_{1}^{-+}(w,w_{1})}{\delta\mB_{-y}}\,G_{0\,+-}(w_{1},w_{2})\,\frac{\delta  S_{1}^{-+}(w_{2},z)}{\delta\mB_{+x}}\,.
\eeqar
For the second term of \eq{C12} we obtain:
\beq\label{C14}
2\,\imath\, S_{xy\,1\,2}\,= \,-\,
G_{0\,+-}(z,w)\,\frac{\delta S_{1}^{-+}(w,w_{1})}{\delta\mB_{-y}}\,G_{0\,+-}(w_{1},w_{2})\,\frac{\delta  S_{1}^{-+}(w_{2},z)}{\delta\mB_{+x}}\,,
\eeq
it cancels with the last term (or term before the last) from the previous expression. So, the contribution we have are represented by the three first terms of \eq{C13} kernel.
We calculate them consequently one by one, the results we obtained are listed below.
\begin{enumerate}
\item 
The first term we have is
\beqar
&2&\imath\, S_{xy\,1\,1\,1}\,= \,-\,\frac{S^{-+}_{2}(w,z)}{\delta \mB_{+x}\delta \mB_{-y}}\,G_{0\,+-}(z,w)\,=\,\frac{1}{2\kappa^2}\,\int d^4 z\,\int d^4 w\,
\delta(x^{+}-w^{+})\delta(y^{-}-z^{-})\delta^{2}(x_{\bot}-w_{\bot})\delta^{2}(y_{\bot}-z_{\bot})\,
\nonumber 
\\
&\,&\D_{ix}^{2}\,\D_{iy}^{2}\,\Le\eta_{jk}\,\D_{jx}\, \D_{ky}\Ra\,
\int ds^{-}\,f(s^{-},w^{-})\int dt^{+}\,f(t^{+},z^{+})\,G_{0\,-j+k}(w,z)\,\D_{ix}^{2}\,\D_{iy}^{2}\,
G_{0\,+-}(t^{+},z^{-},z_{\bot},w^{+},s^{-},w_{\bot})\,=\,
\nonumber \\
&=&\,\D_{ix}^{2}\,\D_{iy}^{2}\,\Le\eta_{jk} \D_{jx}\, \D_{ky}\Ra\,\int\,dp_{+}\,\frac{1}{p_{+}\,-\,\imath \varepsilon}\,
\int dp_{-}\,\frac{1}{p_{-}\,+\,\imath \varepsilon}\,
\int\,\frac{d^{2}p_{\bot}}{(2\pi)^{4}}\,\frac{e^{-\imath\,(p_{i}\,-\,k_{i})(x_{i}\,-\,y_{i})}}{2p_{+}p_{-}\,-\,p_{\bot}^{2}\,+\,\imath \varepsilon}
\nonumber \\
&\,&
\int\,\frac{d^{2}k_{\bot}}{(2\pi)^{2}}\,\frac{1}{2p_{+}p_{-}\,-\,k_{\bot}^{2}\,+\imath \varepsilon}\,=\,
\D_{ix}^{2}\,\D_{iy}^{2}\,\Le\eta_{jk} \D_{jx}\, \D_{ky}\Ra\,
\int\,\frac{d^{2}p_{\bot}}{(2\pi)^{4}}\,\int\,\frac{d^{2}k_{\bot}}{(2\pi)^{2}}\,e^{-\imath\,(p_{i}\,-\,k_{i})(x_{i}\,-\,y_{i})}
\nonumber\\
&\,&
\int dp_{-}\,\frac{1}{p_{-}\,(4p_{-})^{2}}\,
\int\,\frac{dp_{+}}{p_{+}\,-\,\imath \varepsilon}\,
\frac{1}{p_{+}\,-\,p_{\bot}^{2}/2p_{-}\,+\imath \varepsilon}\,\frac{1}{p_{+}\,-\,k_{\bot}^{2}/2p_{-}\,+\imath \varepsilon}\,=\,
\nonumber \\
&=&
\frac{\imath}{2 \pi}\,\D_{ix}^{2}\,\D_{iy}^{2}\,\Le\eta_{jk} \D_{jx}\, \D_{ky}\Ra\,\int \frac{dp_{-}}{p_{-}}\,
\int\,\frac{d^{2}p_{\bot}}{(2\pi)^2}\,\int\,\frac{d^{2}k_{\bot}}{(2\pi)^{2}}\,\frac{e^{-\imath\,(p_{i}\,-\,k_{i})(x_{i}\,-\,y_{i})}}{p_{\bot}^{2}\,k_{\bot}^{2}}\,.
\label{C15}	
\eeqar	
We notice, that except the sign and numericall coefficient, the structure of the expression is the same as the corresponding term of the \eq{C6} effective action  
after it's variation with respect to the reggeized fields and substitution of the $<\varepsilon_{-}\,\varepsilon_{+}>$ correlator directly in the 
expression., that clear from the QFT calculation principles. This correspondence helps to check correctness of the obtained answers. 
We also note, that the non-zero $\gamma$ coefficient of \eq{CE2003} expression provides here an additional $2$ factor. Namely,  for  $\gamma\,=\,0$ case, which is Liparov's 
\cite{LipatovGrav,LipatovGrav01,LipatovGrav02,LipatovGrav03,LipatovGrav04,LipatovGrav05,LipatovGrav06}
result, the overall numerical coefficient in front of \eq{C15} is $1/4$.
\item
Next term we calculate is
\beqar
&2&\imath\, S_{xy\,1\,1\,2}\,= \,\frac{\delta S^{--}(w,w_1)}{\delta \mB_{+x}}\,G_{0\,-+}(w_1,w_2)\,\frac{\delta S^{++}(w_2,z)}{\delta \mB_{-y}}\,G_{0\,+-}(z,w)\,=\,
\int dz^{+}\,\int dw^{+}_{2}\,\int dw^{-}\,\int dw^{+}_{1}\,
\nonumber\\
&\,&\frac{1}{4}\,\D_{i\,x}^{2}\,\D_{i\,y}^{2}\,
\int  \frac{d^{2} p_{\bot}}{(2\pi)^{4}}\,\int  \frac{d^{2} k_{\bot}}{(2\pi)^{4}}\,\Le p_i k_i\Ra^{2}\,e^{e^{-\imath\,(p_{i}\,-\,k_{i})(x_{i}\,-\,y_{i})}}\,
\int\,\frac{dp_{+} dp_{-}}{p^{2}\,+\,\imath\varepsilon}\,\int\,\frac{dk_{+} dk_{-}}{k^{2}\,+\,\imath\varepsilon}\,e^{\imath\,x^{+}\,(p_{+}\,-\,k_{+})\,-\,\imath\,y^{-}\,(p_{-}\,-\,k_{-})}\,
\nonumber \\
&\,&
\int ds^{-}\,\,f(w^{-}_{1},s^{-})\,f(s^{-},w^{-})\int ds^{+}\,f(z^{+},s^{+})\,f(s^{+},w^{+}_{2})\,
e^{-\,\imath\,p_{+}\,z^{+}}\,e^{\imath\,p_{-}\,w^{-}}\,e^{\imath\,k_{+}\,w^{+}_{2}}\,e^{-\,\imath\,k_{-}\,w^{-}_{1}}\,=\,
\nonumber \\
&=&\,
\frac{1}{4}\,\D_{i\,x}^{2}\,\D_{i\,y}^{2}\,\int  \frac{d^{2} p_{\bot}}{(2\pi)^{3}}\,\int  \frac{d^{2} k_{\bot}}{(2\pi)^{3}}\,\Le p_i k_i\Ra^{2}\,
e^{-\imath\,(p_{i}\,-\,k_{i})(x_{i}\,-\,y_{i})}\,
\int\,\frac{dp_{+} dp_{-}}{\Le p^{2}\,+\,\imath\varepsilon\Ra\,\Le p_{+}\,-\,\imath\varepsilon \Ra\,\Le p_{-}\,-\,\imath\varepsilon  \Ra}\,
\nonumber \\
&\,&
\int\,\frac{dk_{+} dk_{-}}{\Le k^{2}\,+\,\imath\varepsilon\Ra\,\Le k_{+}\,-\,\imath\varepsilon \Ra\,\Le k_{-}\,-\,\imath\varepsilon  \Ra}\,
e^{\imath\,x^{+}\,(p_{+}\,-\,k_{+})\,-\,\imath\,y^{-}\,(p_{-}\,-\,k_{-})}\,
\delta(p_{+}\,-\,k_{+})\,\delta(p_{-}\,-\,k_{-})\,=\,
\nonumber \\
&=&\,\frac{1}{4}\,\D_{i\,x}^{2}\,\D_{i\,y}^{2}\,
\int  \frac{d^{2} p_{\bot}}{(2\pi)^{3}}\,\int  \frac{d^{2} k_{\bot}}{(2\pi)^{3}}\,\Le p_i k_i\Ra^{2}\,e^{e^{-\imath\,(p_{i}\,-\,k_{i})(x_{i}\,-\,y_{i})}}\,
\nonumber \\
&\,&
\int\,\frac{dp_{+} dp_{-}}{\Le 2p_{+}\,p_{-}\,-\,p_{\bot}^{2}\,+\,\imath\varepsilon\Ra\,\Le 2p_{+}\,p_{-}\,-\,k_{\bot}^{2}\,+\,\imath\varepsilon \Ra}\,
\frac{1}{\Le p_{+}\,-\,\imath\varepsilon \Ra^{2}\,\Le p_{-}\,-\,\imath\varepsilon  \Ra^{2}}\,.
\label{C16}	
\eeqar	
Here we used the $f$ function representation of the following type:
\beq\label{C17}	
\int dz^{+}\,f(z^{+},s^{+})\,e^{-\imath\,p_{+}\,z^{+}}\,=\,\int_{s^{+}}^{\infty} dz^{+}\,e^{-\imath\,(p_{+}\,-\,\imath\varepsilon)\,z^{+}}\,=\,
-\,\frac{\imath}{p_{+}\,-\,\imath\varepsilon}\,e^{-\imath\,p_{+}\,s^{+}}\,.
\eeq
The resulting positions of the poles in the expression coincide with the result of \cite{LipatovGrav,LipatovGrav01,LipatovGrav02,LipatovGrav03,LipatovGrav04,LipatovGrav05,LipatovGrav06} results. The integral can be calculated enclosing the integration contour in the upper or lower
parts of the integration plane, the answers are the same in both cases, the plane we choose is a plane of complex $p_{+}$ coordinate. We obtain therefore: 
\beqar
&2&\,\imath\, S_{xy\,1\,1\,2}\,= \,\frac{-2\pi\imath}{4}\,\D_{i\,x}^{2}\,\D_{i\,y}^{2}\,
\nonumber \\
&\,&
\int  \frac{d^{2} p_{\bot}}{(2\pi)^{3}}\,\int  \frac{d^{2} k_{\bot}}{(2\pi)^{3}}\,\Le p_i k_i\Ra^{2}\,e^{e^{-\imath\,(p_{i}\,-\,k_{i})(x_{i}\,-\,y_{i})}}\,
\int\,\frac{dp_{-}}{(2p_{-})^{2}\,p_{-}^{2}}\,
\Le \frac{(2p_{-})^3}{p_{\bot}^{4}\,\Le p_{\bot}^{2}-k_{\bot}^{2} \Ra}\,+\,\frac{(2p_{-})^3}{k_{\bot}^{4}\,\Le k_{\bot}^{2}-p_{\bot}^{2} \Ra}\Ra\,=\,
\nonumber \\
&=&\,
\frac{\imath}{4\pi}\,\D_{i\,x}^{2}\,\D_{i\,y}^{2}\,\int \frac{dp_{-}}{p_{-}}\,\int  \frac{d^{2} p_{\bot}}{(2\pi)^{2}}\,\int  \frac{d^{2} k_{\bot}}{(2\pi)^{2}}\,\Le p_i k_i\Ra^{2}\,
e^{-\imath\,(p_{i}\,-\,k_{i})(x_{i}\,-\,y_{i})}\,\frac{p_{\bot}^{2}\,+\,k_{\bot}^{2}}{p_{\bot}^{4}\,k_{\bot}^{4}}\,.
\label{C18}	
\eeqar
In both answers we used \eq{Sca301}-\eq{Sca30101} form of the propagators, the structure of the terms coincides with the amplitude's expression presented in 
\cite{LipatovGrav,LipatovGrav01,LipatovGrav02,LipatovGrav03,LipatovGrav04,LipatovGrav05,LipatovGrav06} paper.

\item
The last terms we need is the following one:
\beqar\label{C19}	
&2&\imath\, S_{xy\,1\,1\,3}\,=\,G_{0\,+-}(z,w)\,\frac{\delta S_{1}^{-+}(w,w_{1})}{\delta\mB_{+x}}\,G_{0\,+-}(w_{1},w_{2})\,\frac{\delta  S_{1}^{-+}(w_{2},z)}{\delta\mB_{-y}}\,=\, 	
\frac{1}{4}\,\D_{i\,x}^{2}\,\D_{i\,y}^{2}\,
\nonumber \\
&\,&
\int  \frac{d^{2} p_{\bot}}{(2\pi)^{3}}\,\int  \frac{d^{2} k_{\bot}}{(2\pi)^{3}}\,p_{\bot}^{2}\,k_{\bot}^{2}\,e^{e^{-\imath\,(p_{i}\,-\,k_{i})(x_{i}\,-\,y_{i})}}\,
\int\,\frac{dp_{+} dp_{-}}{\Le 2p_{+}\,p_{-}\,-\,p_{\bot}^{2}\,+\,\imath\varepsilon\Ra\,\Le 2p_{+}\,p_{-}\,-\,k_{\bot}^{2}\,+\,\imath\varepsilon \Ra}\,.
\frac{1}{\Le p_{+}\,-\,\imath\varepsilon \Ra^{2}\,\Le p_{-}\,-\,\imath\varepsilon  \Ra^{2}}\,
\nonumber
\eeqar	
and it provides almost the same contribution as \eq{C18} answer does:
\beq\label{C20}	
2\,\imath\, S_{xy\,1\,1\,3}\,=\,
\frac{\imath}{4\pi}\,\D_{i\,x}^{2}\,\D_{i\,y}^{2}\,\int \frac{dp_{-}}{p_{-}}\,\int  \frac{d^{2} p_{\bot}}{(2\pi)^{2}}\,\int  \frac{d^{2} k_{\bot}}{(2\pi)^{2}}\,
e^{-\imath\,(p_{i}\,-\,k_{i})(x_{i}\,-\,y_{i})}\,\frac{p_{\bot}^{2}\,+\,k_{\bot}^{2}}{p_{\bot}^{2}\,k_{\bot}^{2}}\,,
\eeq
this terms is absent in the $\gamma\,=\,0$ case of course.

\end{enumerate}

\newpage
\section{ Effective vertices of interaction of reggeized gravitons}\label{AppD}
\renewcommand{\theequation}{D.\arabic{equation}}
\setcounter{equation}{0}
 The interaction term we obtain after the integration in respect to $\epsilon_{\pm i}$ fluctuation has the following form:
\beqar
S_{int}\,&=&-
\frac{1}{4\kappa^{2}}
\int\, d^4 x\,d^4 y\,\Le\D_{i}^{2} \mB_{++}\D_{j}\D_{-}^{-1}B_{--}+\D_{i}^{2} \mB_{--}\mB_{++}\D_{j}\D_{-}^{-1}\Ra\,
G_{0\,- j + k}(x,y)
\Le\D_{k}\D_{+}^{-1}\D_{i}^{2} \mB_{++}\mB_{--}+\D_{i}^{2} \mB_{--}\D_{k}\D_{+}^{-1}B_{++}\Ra\,+\,  
\nonumber \\
&+&
\frac{1}{4\kappa}\,\int\, d^4 x\, \D_{i}^{2} \mB_{++}\, \D_{j}\D_{-}^{-1}B_{--}\,\D_{j}\D_{-}^{-1}\epsilon_{--}\,+\,
\frac{1}{4\kappa}\,\int\, d^4 x\, \D_{i}^{2} \mB_{--}\, \D_{j}\D_{+}^{-1}B_{++}\,\D_{j}\D_{-}^{-1}\epsilon_{++}\,-\,
\nonumber \\
&+&
\frac{1}{4\kappa}\,\int\, d^4 x\, \D_{i}^{2} \mB_{--}\,\mB_{++}\, \D_{j}^{2}\D_{-}^{-1}\D_{-}^{-1}\epsilon_{--}\,+\,
\frac{1}{4\kappa}\,\int\, d^4 x\, \D_{i}^{2} \mB_{++}\,\mB_{--}\, \D_{j}^{2}\D_{+}^{-1}\D_{-}^{-1}\epsilon_{++}\,+\,
\nonumber \\
&+&
\frac{1}{4\kappa}\,\int\, d^4 x\, \D_{i}^{2} \mB_{++}\,\D_{j}^{2}\D_{+}^{-1}\D_{+}^{-1}\mB_{++}\, \epsilon_{--}\,+\,
\frac{1}{4\kappa}\,\int\, d^4 x\, \D_{i}^{2} \mB_{--}\,\D_{j}^{2}\D_{-}^{-1}\D_{-}^{-1}\mB_{--}\, \epsilon_{++}\,+\,
\nonumber \\
&+&\,
\frac{1}{8\kappa}\,\int\, d^4 x\,\D_{i}^{2} \mB_{++}\,\Le \D_{i}\D_{-}^{-1}\mB_{--}\Ra^{2}\,+\,
\frac{1}{8\kappa}\,\int\, d^4 x\,\D_{i}^{2} \mB_{--}\,\Le \D_{i}\D_{+}^{-1}\mB_{++}\Ra^{2}\,+\,
\nonumber \\
&+&
\frac{1}{4\kappa}\,\int\, d^4 x\, \D_{i}^{2} \mB_{--}\,\mB_{++}\, \D_{j}^{2}\D_{-}^{-1}\D_{-}^{-1}\mB_{--}\,+\,
\frac{1}{4\kappa}\,\int\, d^4 x\, \D_{i}^{2} \mB_{++}\,\mB_{--}\, \D_{j}^{2}\D_{-}^{-1}\D_{-}^{-1}\mB_{++}\,.
\label{D1}
\eeqar
Now, after the integrating out the $\varepsilon_{\pm}$ fluctuations, we have correspondingly:
\beqar
S_{int}(\mB)\, & = &-
\frac{1}{4\kappa^{2}}
\int\, d^4 x\,d^4 y\,\Le\D_{i}^{2} \mB_{++}\D_{j}\D_{-}^{-1}B_{--}+\D_{i}^{2} \mB_{--}\mB_{++}\D_{j}\D_{-}^{-1}\Ra\,
G_{0\,- j + k}(x,y)
\Le\D_{k}\D_{+}^{-1}\D_{i}^{2} \mB_{++}\mB_{--}+\D_{i}^{2} \mB_{--}\D_{k}\D_{+}^{-1}B_{++}\Ra\,-\,  
\nonumber \\
&-&\,
\frac{1}{16\kappa^2}\,\int\,d^4 x\,d^4 y\,
\Le \D_{i}^{2}\mB_{++}\,\D_{j}\D_{-}^{-1}\mB_{--}\,\D_{j}\D_{-}^{-1}+\D_{i}^{2} \mB_{++}\,\D_{j}^{2}\D_{+}^{-1}\D_{+}^{-1}\mB_{++}\Ra
G_{\,--\,++}(x,y)\,
\Le \D_{k}\D_{+}^{-1}\D_{i}^{2}\mB_{--}\,\D_{k}\D_{+}^{-1}\mB_{++}\,+\,
\right.
\nonumber \\
&+&
\left.
\D_{i}^{2} \mB_{--}\,\D_{j}^{2}\D_{-}^{-1}\D_{-}^{-1}\mB_{--}\Ra\,-\,
\nonumber \\
&-&
\frac{1}{32\kappa^2}\,\int\,d^4 x\,d^4 y\,
\Le \Le \D_{i}^{2}\mB_{++}\Ra\,\Le \D_{j}\D_{-}^{-1}\mB_{--}\Ra\Ra\,
\Le\D_{j}\D_{-}^{-1}\,G_{\,--\,--}(x,y)\,\D_{k}\D_{-}^{-1}\,\Ra\,
\Le \Le \D_{i}^{2}\mB_{++}\Ra\,\Le \D_{k}\D_{-}^{-1}\mB_{--}\Ra\Ra\,-\,
\nonumber \\
&-&
\frac{1}{32\kappa^2}\,\int\,d^4 x\,d^4 y\,
\Le \Le \D_{i}^{2}\mB_{--}\Ra\,\Le \D_{j}\D_{+}^{-1}\mB_{++}\Ra\Ra\,
\Le\D_{j}\D_{+}^{-1}\,G_{\,++\,++}(x,y)\,\D_{k}\D_{+}^{-1}\,\Ra\,
\Le \Le \D_{i}^{2}\mB_{--}\Ra\,\Le \D_{k}\D_{+}^{-1}\mB_{++}\Ra\Ra\,+\,
\nonumber \\
&+&
\frac{1}{8\kappa}\,\int\, d^4 x\,\D_{i}^{2} \mB_{++}\,\Le \D_{i}\D_{-}^{-1}\mB_{--}\Ra^{2}\,+\,
\frac{1}{8\kappa}\,\int\, d^4 x\,\D_{i}^{2} \mB_{--}\,\Le \D_{i}\D_{+}^{-1}\mB_{++}\Ra^{2}\,.
\label{D2}
\eeqar
The terms represents LO interaction terms between reggeized gravitons and can be used for the calculations of unitary corrections to the amplitudes.

\end{document}